\pdfoutput=1
\documentclass[iop]{emulateapj}
\usepackage{amsmath,amssymb,color,verbatim}
\slugcomment{}

\shorttitle{Probing the physical conditions and star formation processes in the Galactic HII region S305}
\shortauthors{L.~K. Dewangan et al.}
\begin{document}

\title{Probing the physical conditions and star formation processes in the Galactic HII region S305}
\author{L.~K. Dewangan\altaffilmark{1}, Saurabh Sharma\altaffilmark{2}, Rakesh Pandey\altaffilmark{2}, 
S. del Palacio\altaffilmark{3}, D.~K. Ojha\altaffilmark{4}, P. Benaglia\altaffilmark{4}, T. Baug\altaffilmark{5}, and S.~R. Das\altaffilmark{6}}
\email{lokeshd@prl.res.in}
\altaffiltext{1}{Physical Research Laboratory, Navrangpura, Ahmedabad - 380 009, India.}
\altaffiltext{2}{Aryabhatta Research Institute of Observational Sciences, Manora Peak, Nainital 263 002, India.}
\altaffiltext{3}{Instituto Argentino de Radioastronom\'ia, Argentina.}
\altaffiltext{4}{Department of Astronomy and Astrophysics, Tata Institute of Fundamental Research, Homi Bhabha Road, Mumbai 400 005, India.}
\altaffiltext{5}{Kavli Institute for Astronomy and Astrophysics, Peking University, 5 Yiheyuan Road, Haidian District, Beijing 100871, P. R. China.}
\altaffiltext{6}{Indian Institute of Space Science and Technology, Trivandrum 695547, India.}
%

\begin{abstract}
We present multi-scale and multi-wavelength observations of the Galactic H\,{\sc ii} region S305, 
which is excited by massive O8.5V and O9.5V stars. Infrared images reveal an extended sphere-like shell 
(extension $\sim$7.5 pc; at $T_\mathrm{d}$ = 17.5--27~K) enclosing the S305 H\,{\sc ii} region (size $\sim$5.5 pc; age $\sim$1.7 Myr). The extended structure observed in the  {\it Herschel} temperature map indicates that the molecular environment of S305 is heated by the massive O-type stars. Regularly spaced molecular condensations and dust clumps are investigated toward the edges of the infrared shell, 
where the PAH and H$_{2}$ emission is also observed. 
The molecular line data show a signature of an expanding shell of molecular gas in S305. GMRT 610 and 1280 MHz continuum maps reveal overdensities of the ionized emission distributed around two O-type stars, which are 
surrounded by the horseshoe envelope (extension $\sim$2.3 pc). A molecular gas deficient region/cavity is identified toward the center of the horseshoe envelope, which 
is well traced with PAH, H$_{2}$, molecular, and dust emission.
The edges of the infrared shell are found to be located in the front of the horseshoe envelope. 
All these outcomes provide the observational evidence of the feedback of O-type stars in S305. 
Moreover, non-thermal radio emission is detected in S305 with an average spectral index $\alpha \sim-0.45$.~The variations in $\alpha$, ranging from $-1.1$ to 1.3, are explained due to soft synchrotron 
emission and either optically-thicker thermal emission at high frequencies or a suppression of 
the low-frequency emission by the Razin-Tsytovich effect.
\end{abstract}
%
\section{Introduction}
\label{sec:intro}
%
Massive OB-stars ($\geq 8$~M$_{\odot}$) have a profound impact on their immediate environment through the radiative and mechanical energy they inject from their formation phase until their death. 
Hence, these stars have the ability to trigger the birth of young stellar objects (YSOs) and young massive stars \citep{elmegreen77,elmegreen98,deharveng05,deharveng10}. However, understanding the formation mechanisms of massive OB-stars and their feedback process are still far from complete \citep{zinnecker07,tan14}. 
In the literature, one can find two major triggered star formation scenarios concerning the expansion of the H\,{\sc ii} regions, which are the radiation-driven implosion \citep[RDI; see][]{bertoldi89,lefloch94} and the ``collect and collapse" \citep[see][]{elmegreen77,whitworth94,dale07}. In the RDI model, the expansion of an H\,{\sc ii} region initiates the instability and helps in the collapse of a pre-existing dense region in the molecular cloud. In the ``collect and collapse" scenario, the expansion of an H\,{\sc ii} region collects a massive and dense shell of cool neutral material between the ionization and the shock fronts, where star formation is initiated when this material becomes gravitationally unstable; the formation of massive stars is expected in this particular scenario \citep[e.g.,][]{deharveng10}. These theoretical models are observationally difficult to assess. One can find the details of the observational challenges and difficulties for searching true sites of triggered star formation in \citet{dale15}. Some indirect evidences of triggering have been reported in the literature \citep[e.g.,][]{deharveng05,deharveng10}.

In this paper, we have selected the Galactic H\,{\sc ii} region Sh 2-305 (hereafter, S305), which is located at a distance of $\sim 3.7$~kpc \citep{pandey20}. Based on the previously published mid-infrared (MIR) images, S305 can be considered as a good example of a MIR bubble \citep[see Figure~1 in][]{pandey20}. Approximately at its center, S305 harbors two spectroscopically identified massive O-type stars \citep[O8.5V: VM4 and O9.5V: VM2;][]{vogt75,chini84,russeil95,pandey20}. 
In the direction of S305, the ionized gas was traced in a radial velocity of $\sim 35.4$~km~s$^{-1}$ \citep[e.g.,][]{balser11,hou14}.
Using the James Clerk Maxwell Telescope (JCMT) $^{12}$CO(2--1) and $^{13}$CO(2--1) line data, \citet{azimlu11} studied the properties of molecular clouds associated 
with 10 H\,{\sc ii} regions including S305, and the molecular gas toward S305 was examined in a velocity range of about 42--49 km s$^{-1}$ (see Tables~2 and~3 in their paper). 
Using optical and infrared (IR) photometric data, \citet{pandey20} identified and studied the distribution of 116 YSOs in S305 in an area of $\sim 18.5\arcmin \times 18.5\arcmin$. These authors observed at least three stellar sub-clusterings in S305 (see Figure~3 in their paper). 
One of the clusters is detected toward the center of S305, which spatially coincides with the previously known young open star cluster ``Mayer~3" \citep{vogt75,chini84,russeil95}. The site S305 was mapped in the JCMT SCUBA2 450 and 850~$\mu$m continuum maps, and 
several clumps were observed in S305 \citep[see Figure~5 in][]{sreenilayam14}. 
Using the {\it Herschel} column density map (resolution $\sim 12\arcsec$), at least 25 clumps (mass range $\sim$35--1565 M$_{\odot}$) have been reported in S305, and star formation activities are traced at some of these clumps \citep[e.g.,][]{pandey20}. 
\citet{pandey20} suggested that the two massive O-type stars might have stimulated the birth of young stars in S305. 

The present paper aims to understand the physical processes operational in S305, including the impact of massive O-type stars on their surrounding molecular environment as well as ongoing and past star formation activity.  
In this connection, a multi-scale and multi-wavelength observational approach has been employed. We study the site S305 using high angular resolution radio continuum maps at 610 and 1280~MHz observed with the Giant Metrewave 
Radio Telescope (GMRT) facility. These maps enable us to examine the inner structure of the S305 H\,{\sc ii} region as well as the nature of radio continuum emission in S305.  
Our work also uses publicly available optical H$\alpha$, IR, sub-millimeter, molecular line, and radio continuum data sets of 
S305 (see Section~\ref{sec:obser}), allowing to infer the velocity structure of molecular gas and the physical conditions in the target site. 

We present the details of the adopted observational data sets in Section~\ref{sec:obser}. In Section~\ref{sec:data}, the observational findings and derived parameters are presented.
We discuss the observational results and ongoing physical processes in Section~\ref{sec:disc}.
Finally, our conclusions are summarized in Section~\ref{sec:conc}.
%
\section{Data and analysis}
\label{sec:obser}
%
We select a target region centered at $l = 233.75\degr$; $b = -0.19\degr$ of size $\sim 18.7\arcmin \times 18.7\arcmin$ (20.1 pc $\times$ 20.1 pc at a distance of 3.7~kpc). 
\subsection{Radio Continuum Observations}
\label{sec:gmrt}
The present paper uses new radio continuum observations at 610 and 1280 MHz taken with the GMRT facility on 2017 October 30 \& 31, and 2017 November 12 (Proposal Code: 33$\_$065; PI: Rakesh Pandey).   
We reduced the GMRT data using the Astronomical Image Processing System (AIPS) package \citep[see][for the detailed reduction procedures]{mallick12,mallick13}. We utilized the VLA calibrators 3C147 and 3C286 as flux calibrators, while 0744$-$064 was used as a phase calibrator. 
The bad data from the \textit{uv} data were flagged out by multiple rounds of flagging using the {\sc tvflg} task of the AIPS. 
After performing several rounds of `self-calibration', we generated the final maps at 610 and 1280 MHz with synthesized beams 
of $8''.3 \times 5''.8$ and $4''.6 \times 3''.0$, respectively. 
Furthermore, we also convolved both GMRT maps to a common resolution of $10'' \times 10''$ for further analysis. The antenna temperature of the sources can be increased in the direction of the Galactic plane due to the Galactic background emission. The Galactic background contamination is expected to be significant only in the 610 MHz band. Hence, we apply the system temperature correction to the GMRT 610 MHz data before performing the analysis; a more detailed description of this correction is given in \citet[][and references therein]{baug15}. 
We determined the final rms sensitivities of the maps at 610 and 1280 MHz to be $\sim$0.46 and $\sim$0.74~mJy beam$^{-1}$, respectively.
\subsection{Archival Data}
\label{sec:archive}
We employed several existing multi-wavelength data sets obtained from various large-scale
surveys; namely, the NRAO VLA Sky Survey \citep[NVSS; $\lambda$ =21 cm; resolution $\sim$46$''$;][]{condon98}, 
the FOREST Unbiased Galactic plane Imaging survey with the Nobeyama 45-m telescope \citep[FUGIN; $^{12}$CO and $^{13}$CO; resolution $\sim$20$''$;][]{umemoto17}, 
the JCMT SCUBA-2 Guaranteed Time projects (ID: M11BGT01; $\lambda$ =850 $\mu$m; resolution $\sim$14$''$.4; PI: Wayne S. Holland), 
the {\it Herschel} Infrared Galactic Plane Survey \citep[Hi-GAL; $\lambda$ =70--500 $\mu$m; resolution $\sim$5.8--46$''$;][]{molinari10}, 
the Warm-{\it Spitzer} GLIMPSE360 Survey \citep[$\lambda$ =3.6 and 4.5 $\mu$m; resolution $\sim$2$''$;][]{benjamin03,whitney11}, and 
the AAO/UKST SuperCOSMOS H-alpha Survey \citep[SHS; $\lambda$ =0.6563 $\mu$m; resolution $\sim$1$''$;][]{parker05}.

The $^{12}$CO(J =1$-$0) and $^{13}$CO(J =1$-$0) line data obtained from the FUGIN survey are calibrated in main beam temperature \citep[$T_\mathrm{mb}$, see][]{umemoto17}. 
The typical rms noise level\footnote[1]{https://nro-fugin.github.io/status/} ($T_\mathrm{mb}$) is $\sim$1.5~K and $\sim$0.7~K for $^{12}$CO and $^{13}$CO lines, respectively \citep{umemoto17}. To improve the sensitivity, both FUGIN molecular line data cubes are smoothened with a Gaussian function with a full-width at half-maximum of 3 pixels. We have also used the {\it Herschel} temperature and column density ($N(\mathrm H_2)$) maps (resolution $\sim$12$''$) of our selected target area. These maps\footnote[2]{http://www.astro.cardiff.ac.uk/research/ViaLactea/} were generated for the {\it EU-funded ViaLactea project} \citep{molinari10b} using the Bayesian {\it PPMAP} method \citep{marsh15,marsh17}, which was applied on the {\it Herschel} images at wavelengths of 70, 160, 250, 350 and 500 $\mu$m.

Previously, \citet{navarete15} carried out a survey of extended H$_{2}$ emission from massive YSOs, which includes a source G233.8306-00.1803 (ID \#198; see their paper for more details) in the direction of our selected target site. This published narrow-band H$_{2}$ ($\nu$ = $1-0$ S(1) at $\lambda$ = 2.122 $\mu$m ($\Delta \lambda =0.032\, \mu \rm m$)) image (resolution $\sim 0.7''$--$0.9''$) was restricted to a small area of S305 \citep[see source ID \#198 or G233.8306-00.1803 in][]{navarete15}. 
In this paper, we examined the H$_{2}$ image for a larger area ($\sim 7\farcm2 \times 7\farcm2$) around S305, which was observed using the Canada-France-Hawaii Telescope \citep[CFHT, see][]{navarete15}. We also obtained K-band continuum ($\lambda$ = 2.218 $\mu$m; $\Delta \lambda =0.033\, \mu \rm m$) image to produce the final continuum-subtracted H$_{2}$ map \citep[from][]{navarete15}.  
%
\section{Results}
\label{sec:data}
%
%
\subsection{Morphology of S305}
\label{sec:morph}
To unearth the obscured morphology of S305, we examine the distribution of the warm dust emission, molecular gas, and ionized emission. In general, mid-IR images and radio continuum maps are good tracers of warm dust emission and ionized gas, respectively. Molecular line data are employed to study the distribution of cold molecular gas in a given molecular cloud. 
\subsubsection{Infrared shell and horseshoe envelope}
\label{sec:morph1}
\begin{figure*}
\epsscale{0.8}
\plotone{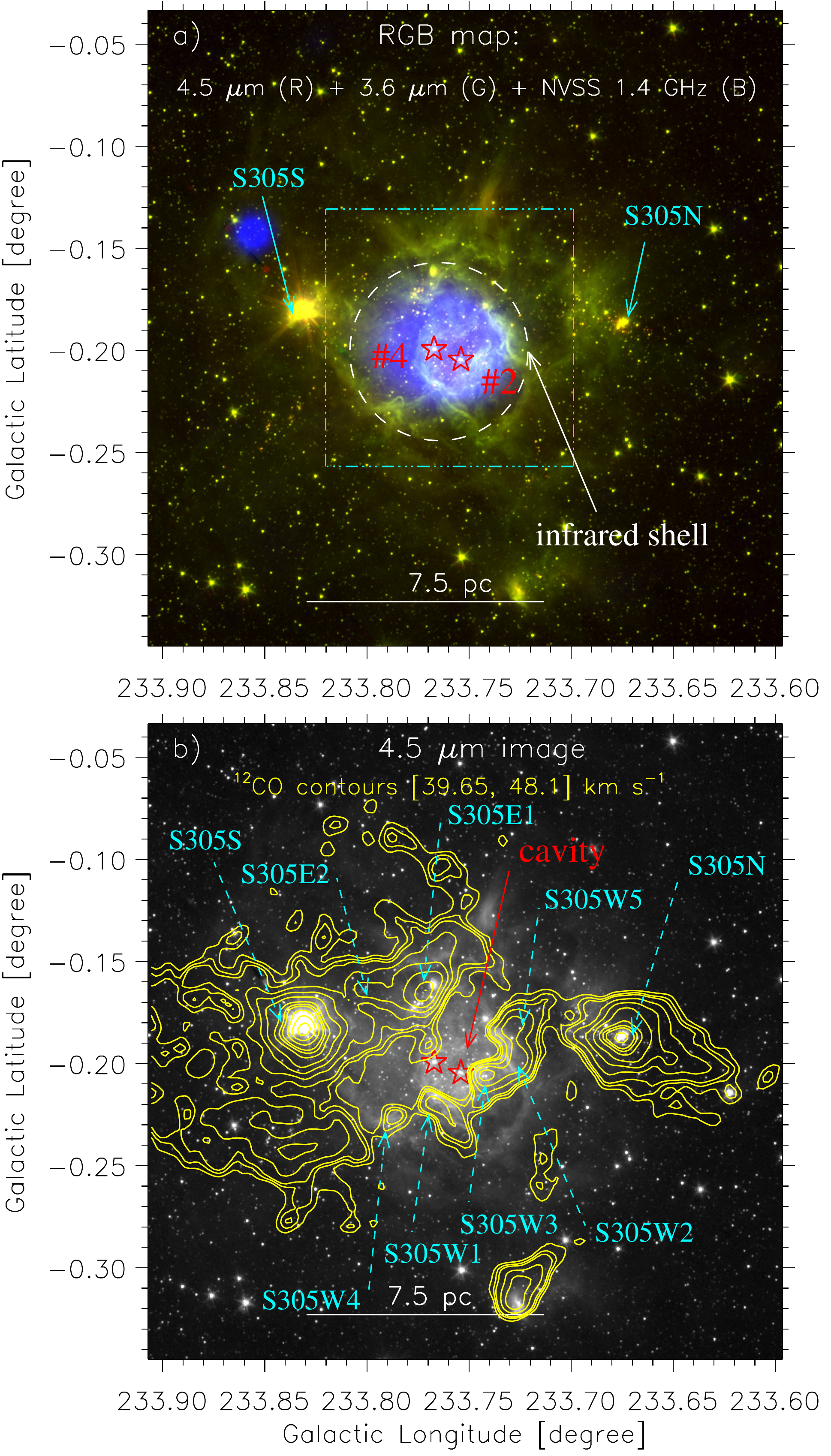}
\caption{a) Three-color composite image ({\it red}, {\it Spitzer} 4.5 $\mu$m; {\it green}, {\it Spitzer} 3.6 $\mu$m; 
{\it blue}, NVSS 1.4 GHz) of the region around S305 (selected area $\sim$18$'$.7 $\times$ 18$'$.7 centered at {\it l} = 233$\degr$.75; {\it b} = $-$0$\degr$.19). An extended shell-like feature is highlighted by a big broken circle (in white). 
A dotted-dashed box (in cyan) encompasses the area shown in Figure~\ref{vfg1}a. 
b) Overlay of the FUGIN $^{12}$CO(J =1$-$0) gas on the {\it Spitzer} 4.5 $\mu$m image. 
The $^{12}$CO integrated velocity range is from 
39.65 to 48.1 km s$^{-1}$ (1$\sigma$ $\sim$2.2 K km s$^{-1}$). The $^{12}$CO contours (in yellow) are 19.5, 22.5, 25, 32, 40, 50, 63, 70, 75, 85, 95, and 105 K km s$^{-1}$. A cavity area is indicated by an arrow (in red). 
Several SCUBA2 850 $\mu$m continuum clumps \citep[from][]{sreenilayam14} are labeled in the figure (see Figure~5 in their paper). 
In each panel, the positions of previously known two massive O9.5V (\#2) and O8.5V (\#4) stars are marked by stars, and a scale bar referring to 7.5 pc (at a distance of 3.7 kpc) is shown.}
\label{ufig1}
\end{figure*}

In Figure~\ref{ufig1}a, we display a three-color composite image made using {\it Spitzer} 4.5 $\mu$m in red, 
{\it Spitzer} 3.6 $\mu$m in green, and NVSS 1.4 GHz in blue. Both {\it Spitzer} images reveal an extended IR shell (extension $\sim$7.5 pc), which has a sphere-like 
morphology (see a dashed circle in Figure~\ref{ufig1}a and arrows in Figure~\ref{ufig1}b). The IR shell is revealed as a most prominent feature of S305, and 
encloses the NVSS radio continuum emission. It is a very similar structure as observed in the triggered star-forming site Sh 2-235 \citep[e.g.,][]{dewangan16}. 
The positions of the previously known massive O8.5V (\#4) and O9.5V (\#2) stars are found approximately at the center of the IR shell. 
To obtain a new insight in the observed morphology of S305, it is essential to examine the molecular gas associated with different parts of S305. In this paper, we employed the FUGIN $^{12}$CO(J =1$-$0) and $^{13}$CO(J =1$-$0) line data to explore the distribution of molecular gas in our selected target area around S305. 
Based on the examination of the average profiles of $^{12}$CO and $^{13}$CO (not shown here), the molecular gas in S305 is studied in a velocity range of [39.65, 48.1]~km~s$^{-1}$. 
Figure~\ref{ufig1}b presents the overlay of the contours of the FUGIN $^{12}$CO intensity map (moment-0) on the {\it Spitzer} 4.5 $\mu$m image (see also Section~\ref{sec:gas}). 
In Figure~\ref{ufig1}b, we also labeled several previously reported dust condensations (e.g., S305N, S305S, S305E1, S305E2, S305W1, S305W2, S305W3, S305W4, and S305W5), which were reported by \citet{sreenilayam14} (see Figure~5 in their paper). We find the presence of molecular gas toward all the dust condensations reported by \citet{sreenilayam14}. 
All these condensations (except S305N) appear to be seen toward the edges of the sphere-like IR shell. 
Additionally, the distribution of molecular gas also shows the existence of an intensity/gas deficient region in S305 
toward the positions of massive O-type stars.  

\begin{figure*}
\epsscale{0.8}
\plotone{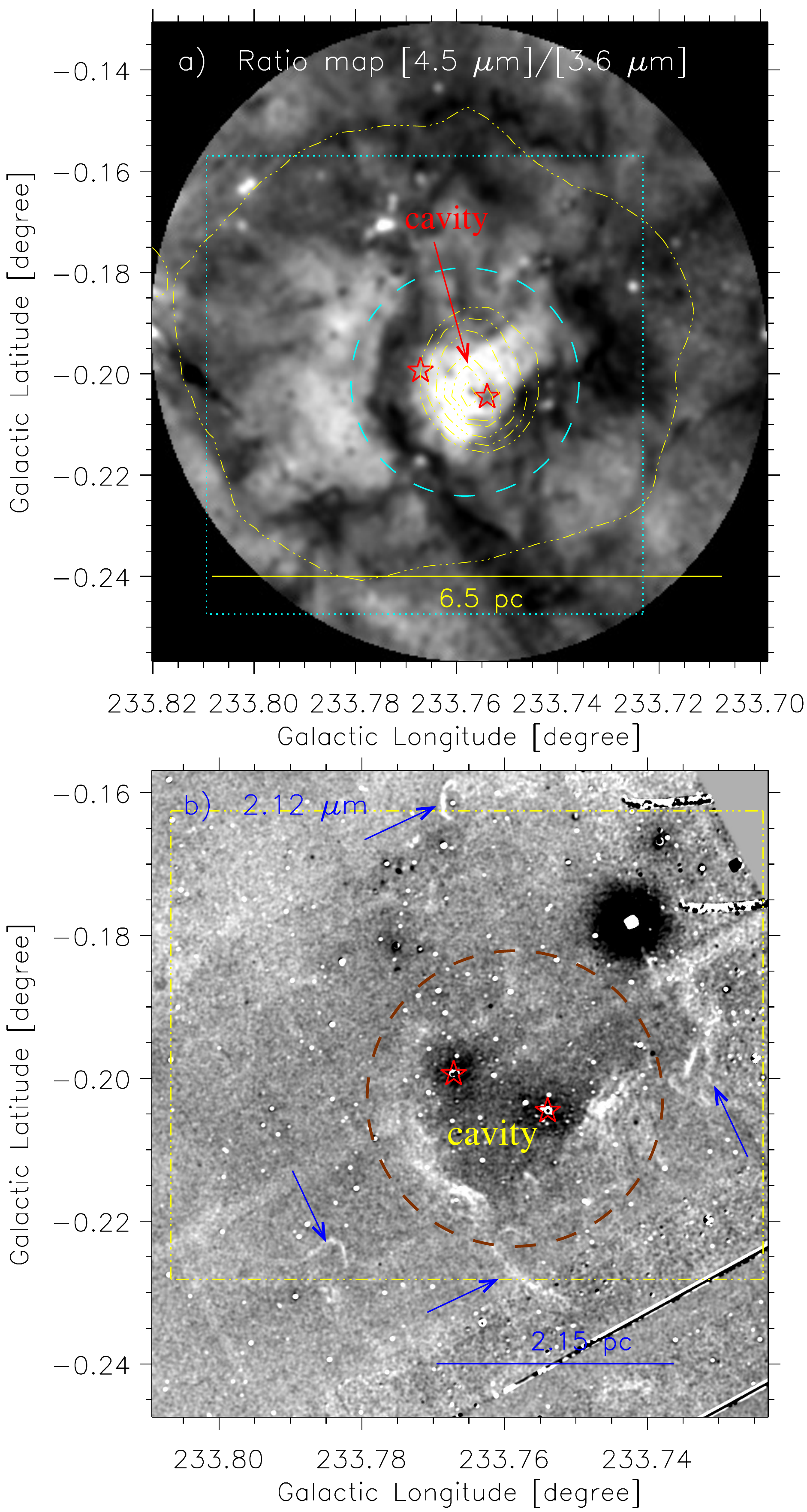}
\caption{a) Overlay of the NVSS radio continuum contours (in yellow) on the {\it Spitzer} ratio map of 4.5 $\mu$m/3.6 $\mu$m emission of an area highlighted by a dotted-dashed 
box in Figure~\ref{ufig1}a. The NVSS contours are shown with the levels of 2.3, 100, 110, 120, 130, 135, and 140 mJy beam$^{-1}$. 
A dotted box (in cyan) encompasses the area shown in Figure~\ref{vfg1}b. 
b) Continuum-subtracted H$_{2}$ image at 2.12 $\mu$m (in gray-scale) of an area highlighted by a dotted 
box in Figure~\ref{vfg1}a. Arrows and a dashed circle indicate the detections of the H$_{2}$ emission. A dotted-dashed box (in yellow) encompasses the area shown in 
Figures~\ref{lfig1}a,~\ref{lfig1}b,~\ref{lfig1}c, and~\ref{lfig1}d. The ratio map and the H$_{2}$ image are smoothened using a Gaussian function with radius of four pixels. 
In each panel, the positions of previously known two massive O9.5V (\#2) and O8.5V (\#4) stars are marked by stars, and 
a dashed circle highlights the immediate surroundings of massive stars (see text for more details).}
\label{vfg1}
\end{figure*}

Figure~\ref{vfg1}a shows the {\it Spitzer} ratio map of 4.5 $\mu$m/3.6 $\mu$m emission in the direction of the IR shell \citep[see also Figure~15 in][]{pandey20} with overlaid NVSS radio continuum emission contours (at 2.3, 100, 110, 120, 130, 135, and 140 mJy beam$^{-1}$). The outermost NVSS radio contour (at 2.3 mJy beam$^{-1}$) follows the edges of the IR shell, while the inner NVSS contours are seen toward the positions of two massive O-type stars. A dashed circle is marked to highlight the immediate surroundings of massive stars. A detailed explanation of the {\it Spitzer} ratio map is given in \citet{pandey20}. In Figure~\ref{vfg1}a, the bright regions hint the presence of the Br$\alpha$ emission at 4.05 $\mu$m, which is observed toward the positions of two massive O-type stars and the peak of the NVSS emission within the IR shell.  
Furthermore, the black or dark gray regions suggest the existence of the polycyclic aromatic hydrocarbon (PAH) emission at 3.3 $\mu$m. Overall, the distribution of the PAH emission traces the edges of the IR shell, and is also found toward the center of the IR shell (see a broken circle in Figure~\ref{vfg1}a). 
Figure~\ref{vfg1}b presents a continuum-subtracted 2.12 $\mu$m H$_{2}$ image of S305, revealing the detections of the H$_{2}$ emission at the periphery of the IR shell as well as toward the center of the IR shell. 
The H$_{2}$ emission spatially coincides with the PAH emission traced in the ratio map (see a dashed circle in Figures~\ref{vfg1}a and~\ref{vfg1}b).
Based on the observed morphology of the H$_{2}$ features, it seems that their origin is due to UV fluorescence \citep[see also][]{dewangan15}.  
The detection of the H$_{2}$, molecular gas, and PAH features surrounding the ionized emission suggests the existence of the photon dominant 
region in S305. The ionizing feedback of two massive O-type stars seems to heat the dust which could be the origin of the PAH, shocked tracer H$_{2}$, and warm dust emission toward the edges of the IR shell (see Section~\ref{sec:dmorph1} for a quantitative estimation). It also includes the absence of molecular gas toward the positions of two massive O-type stars. Hence, all these results suggest strong stellar feedback from the massive O-type 
stars in S305. 

\begin{figure*}
\epsscale{1.1}
\plotone{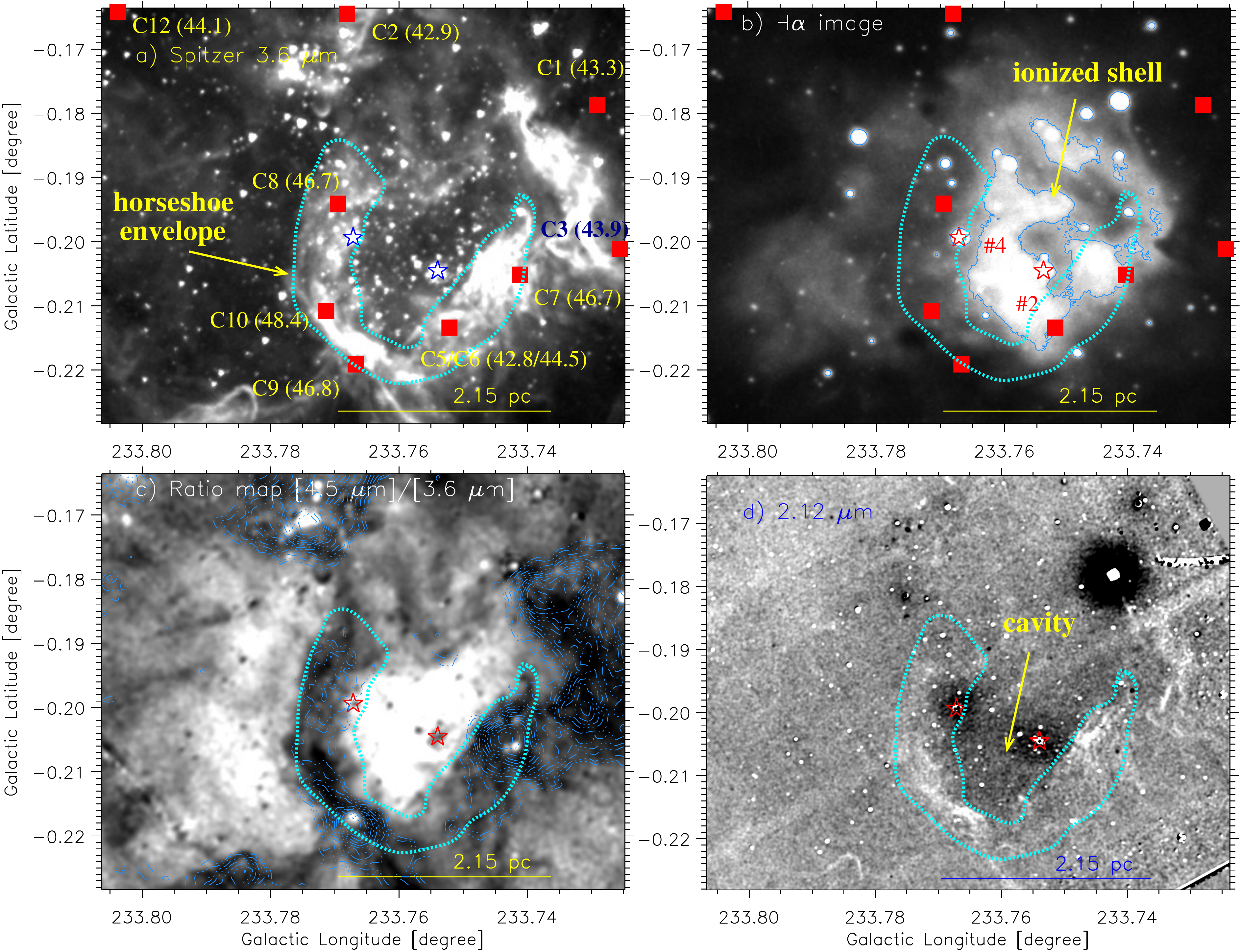}
\caption{a) Gray-scale {\it Spitzer} 3.6 $\mu$m image of an area highlighted by a broken box in 
Figure~\ref{vfg1}b. 
The positions of observed molecular clumps \citep[from][]{azimlu11} are highlighted by filled squares, which 
are also labeled in the figure. The radial velocity of the molecular gas (in km s$^{-1}$) toward each clump \citep[from][]{azimlu11} is also indicated in the figure.
b) Gray-scale H$\alpha$ image of the same area as shown in in Figure~\ref{lfig1}a. 
A solid contour (in dodger blue) shows an ionized shell traced in the H$\alpha$ image. 
Filled squares are the same as shown in Figure~\ref{lfig1}a. 
c) The panel shows the {\it Spitzer} ratio map of 4.5 $\mu$m/3.6 $\mu$m 
emission (see also Figure~\ref{vfg1}a). The ratio map is also overlaid with the 
contours (in dodger blue) of the JCMT SCUBA2 continuum emission at 850 $\mu$m (see also Figure~\ref{fig3}b). 
d) Continuum-subtracted H$_{2}$ image at 2.12 $\mu$m (in gray-scale; see also 
Figure~\ref{vfg1}b). An arrow shows the gas and dust deficient region, which is referred to as cavity. 
In each panel, the positions of previously known two massive O9.5V (\#2) and O8.5V (\#4) stars 
are marked by stars, and a scale bar referring to 2.15 pc (at a distance of 3.7 kpc) is marked. 
In all panels, a dotted curve (in cyan) indicates walls of a cavity.} 
\label{lfig1}
\end{figure*}

We further examine the immediate surrounding environment of the two massive O-type stars in Figure~\ref{lfig1}. Figures~\ref{lfig1}a,~\ref{lfig1}b,~\ref{lfig1}c and~\ref{lfig1}d display the {\it Spitzer} 3.6 $\mu$m, H$\alpha$, {\it Spitzer} ratio map, 
and continuum-subtracted H$_{2}$ images, respectively. 
The ratio map and the H$_{2}$ image are shown here only for comparison purpose (see also Figures~\ref{vfg1}a and~\ref{vfg1}b). 
In Figure~\ref{lfig1}a, the {\it Spitzer} 3.6 $\mu$m traces a horseshoe envelope-like feature (extension $\sim$2.3 pc), which is indicated by a broken cyan curve in the figure. We identify an ionized shell-like feature (extension $\sim$2.15 pc) in the H$\alpha$ image (see Figure~\ref{lfig1}b), which is highlighted by a solid contour in Figure~\ref{lfig1}b. 
This ionized shell appears to be located toward the center of the horseshoe envelope (see a broken cyan curve in Figure~\ref{lfig1}b). 
No $^{12}$CO emission is detected toward the positions of two massive O-type stars or the center of the horseshoe envelope (see Figure~\ref{ufig1}b). The {\it Spitzer} ratio map is also overlaid with the JCMT SCUBA2 850 $\mu$m continuum contours in Figure~\ref{lfig1}c. The horseshoe envelope is also associated with the H$_{2}$, PAH, and dust continuum emission (see a broken cyan curve in Figures~\ref{lfig1}c and~\ref{lfig1}d). In the direction of the horseshoe envelope, these maps also indicate the existence of a cavity, which is filled only with the ionized emission. In other words, the cavity containing the two massive O-type stars is surrounded by dust, CO molecular, H$_{2}$, and PAH emission.
Previously, a very similar configuration was also reported in the sites Sh 2-235 \citep[e.g.,][]{dewangan11,dewangan16} and Sh 2-237 \citep{dewangan17}. 

In Figure~\ref{lfig1}a, we also marked the positions of at least nine molecular clumps \citep[i.e., C1, C2, C3, C5/C6, C7, C8, C9, C10, and C12;][]{azimlu11}. 
The radial velocity of $^{13}$CO(2--1) toward each molecular clump (except C12) is also marked in 
Figure~\ref{lfig1}a (see filled squares). In the case of the molecular clump C12, the radial velocity 
of $^{12}$CO(2--1) is adopted \citep[see][]{azimlu11}. 
Note that the clumps C5 and C6 have the same position with different radial velocities. 
Four molecular clumps (e.g., C1, C2, C3, and C12) are distributed away from the horseshoe envelope.
All these clumps show a velocity range of [42.9, 44.1] km s$^{-1}$.
Other five molecular clumps (e.g., C5/C6, C7, C8, C9, and C10) are seen in the direction of the horseshoe envelope surrounding two massive O-type stars. 
The molecular clumps C7, C8, C9, and C10 are traced in a velocity range of [46.5, 48.5] km s$^{-1}$. 
However, the molecular clump C5/C6 is depicted with a velocity of 42.8/44.5 km s$^{-1}$, which is very similar to that of 
four molecular clumps (i.e., C1, C2, C3, and C12). 
All these findings together hint that the molecular clumps associated with the horseshoe envelope (i.e., C7, C8, C9, and C10) are redshifted compared to other molecular clumps (i.e., C1, C2, C3, C5/C6, and C12). Assuming an expansion model, the molecular clumps associated with the edges of the extended IR shell seem to be located in the front of 
the horseshoe envelope. We have further examined this argument in Section~\ref{sec:gas}.
\subsubsection{Filament, {\it Herschel} clumps, and YSOs}
\label{sec:dmorph1}
\begin{figure*}
\epsscale{0.75}
\plotone{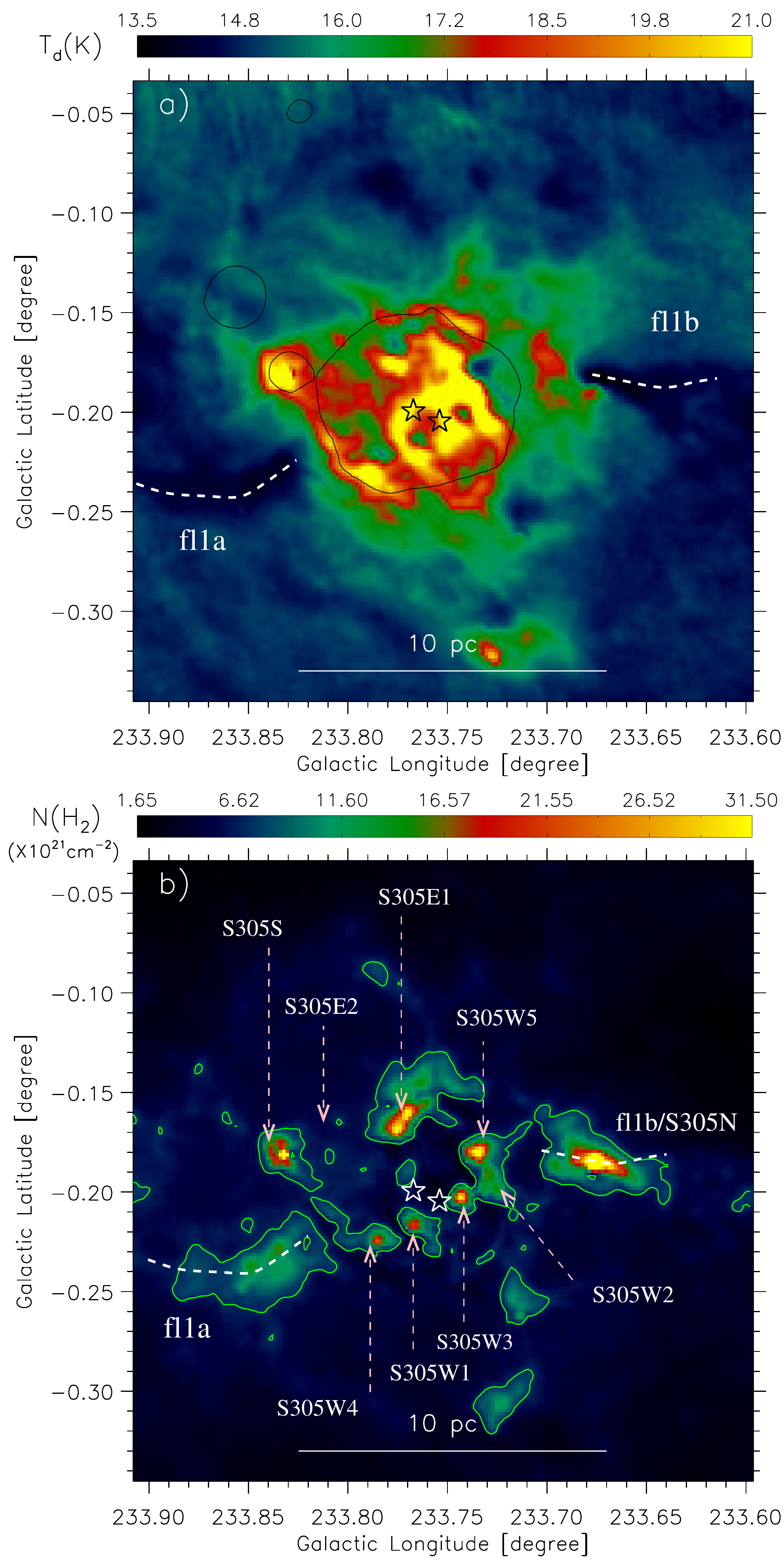}
\caption{a) Overlay of the NVSS radio continuum contour (at 2.3 mJy beam$^{-1}$) on the {\it Herschel} temperature 
map of the selected region around S305. 
b) {\it Herschel} column density ($N(\mathrm H_2)$) map. The $N(\mathrm H_2)$ contour (in green) is 
also shown with a level of 6.85 $\times$ 10$^{21}$ cm$^{-2}$. 
Several SCUBA2 850 $\mu$m continuum clumps \citep[from][]{sreenilayam14} are labeled in the figure (see also Figure~\ref{ufig1}b). 
In both panels, stars show the positions of two massive O-type stars, and two broken curves (in white) highlight both ends of an elongated filament (i.e., fl1a and fl1b). In each panel, the scale bar referring to 10 pc (at a distance of 3.7 kpc) is shown.} 
\label{fig2}
\end{figure*}

Figure~\ref{fig2}a displays the {\it Herschel} temperature map (resolution $\sim$12$''$) of S305, which is also 
superimposed with the NVSS radio continuum emission contour at 2.3 mJy beam$^{-1}$. 
The {\it Herschel} temperature map reveals the extended warm dust emission structures 
(including the IR shell and horseshoe envelope) at $T_\mathrm{d}$ = 17.5--27~K, indicating that massive O-type stars have heated the dust. 
We also marked the relatively cold regions (at $T_\mathrm{d}$ = 13.5--14.2~K) in the temperature map by arrows (i.e., fl1a and fl1b).
It seems that the IR shell containing two massive O-type stars is located at the 
central part of an elongated filament, and its both ends are traced with cold 
regions (see labels fl1a and fl1b in Figure~\ref{fig2}a). 
In Figure~\ref{fig2}b, we present the {\it Herschel} column density ($N(\mathrm H_2)$) 
map (resolution $\sim$12$''$) overlaid with the $N(\mathrm H_2)$ contour (in green) at 6.85 $\times$ 10$^{21}$ cm$^{-2}$ \citep[see also Figure~12a in][]{pandey20}. 
In the {\it Herschel} column density map, materials with high column densities are traced toward the IR shell, 
horseshoe envelope, and obscured filament (see ``fl1a" and ``fl1b" in Figure~\ref{fig2}b). 
We also labeled different dust condensations reported by \citet{sreenilayam14} (see also Figure~\ref{ufig1}b). 

\begin{figure*}
\epsscale{0.9}
\plotone{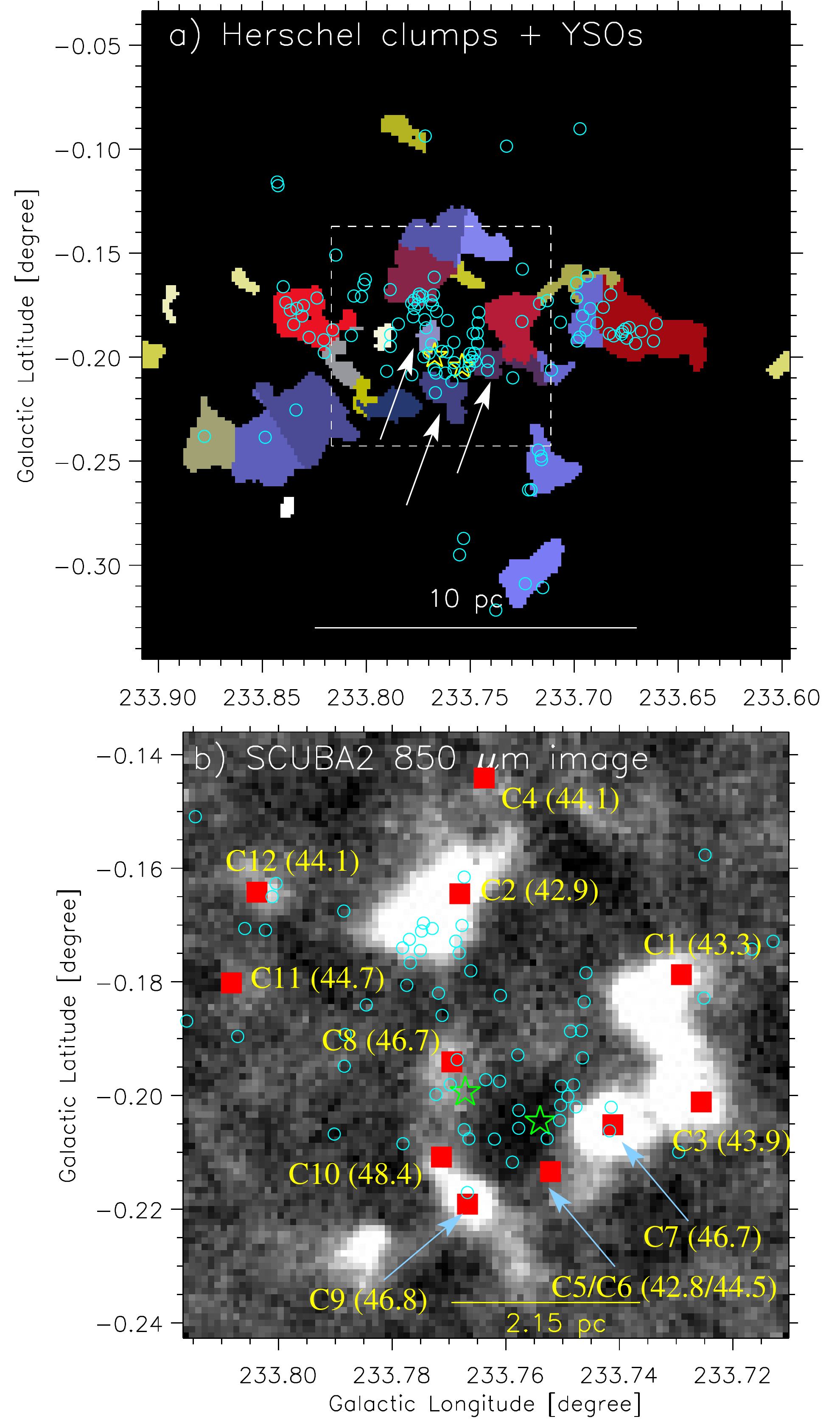}
\caption{a) Distribution of YSOs (see cyan circles) against the boundaries of the {\it Herschel} clumps. Different colors are utilized to distinguish the clumps. All these results are taken from \citet{pandey20} (see Figures~12 and~13 in their paper). Arrows indicate the locations of the {\it Herschel} clumps seen toward the horseshoe envelope. b) JCMT SCUBA2 continuum emission map at 850 $\mu$m of an area highlighted by a dashed box in Figure~\ref{fig3}a. 
The positions of 12 molecular clumps \citep[from][]{azimlu11} are highlighted by filled squares, which are also labeled in the figure. 
Circles show the position of YSOs. 
The radial velocity of the molecular gas toward each clump \citep[from][]{azimlu11} is also indicated in the figure (see also Figure~\ref{lfig1}a).}
\label{fig3}
\end{figure*}

Using the {\it Herschel} column density map, \citet{pandey20} identified several {\it Herschel} clumps in S305, 
and their boundaries are shown in Figure~\ref{fig3}a (see Table~8 in their paper for more details). 
The {\it Herschel} clumps are found toward the IR shell, the filament, and the horseshoe envelope. 
In the direction of the horseshoe envelope, three {\it Herschel} clumps \citep[mass range $\sim$85--320 M$_{\odot}$; see also clump IDs 6, 14, and 16 in Table~8 in][]{pandey20} are observed in the {\it Herschel} column density map (see arrows in Figure~\ref{fig3}a). 
We also adopted the positions of YSOs in S305 \citep[from][]{pandey20}, which are marked by circles in Figure~\ref{fig3}a. Figure~\ref{fig3}a helps us to trace star formation activity toward some {\it Herschel} clumps \citep[see][for more details]{pandey20}. 
In general, \citet{evans09} reported the average ages of Class~I and Class~II YSOs to be $\sim$0.44 Myr and $\sim$1--3 Myr, respectively.  

Using the JCMT SCUBA2 continuum emission map at 850 $\mu$m, Figure~\ref{fig3}b displays a zoomed-in view of S305. 
In Figure~\ref{fig3}b, we marked the positions of YSOs (see open circles) and the positions of 12 molecular clumps \citep[by filled squares; see][]{azimlu11} (see also Figure~\ref{lfig1}a and Section~\ref{sec:morph1}). 
The locations of the condensations seen in the JCMT SCUBA2 map at 850 $\mu$m are in agreement with the {\it Herschel} clumps (see Figure~\ref{fig3}a). 
Noticeable YSOs are seen around both massive O-type stars in the direction of the horseshoe envelope. 
\subsection{Kinematics of molecular gas}
\label{sec:gas}
\begin{figure*}
\epsscale{0.75}
\plotone{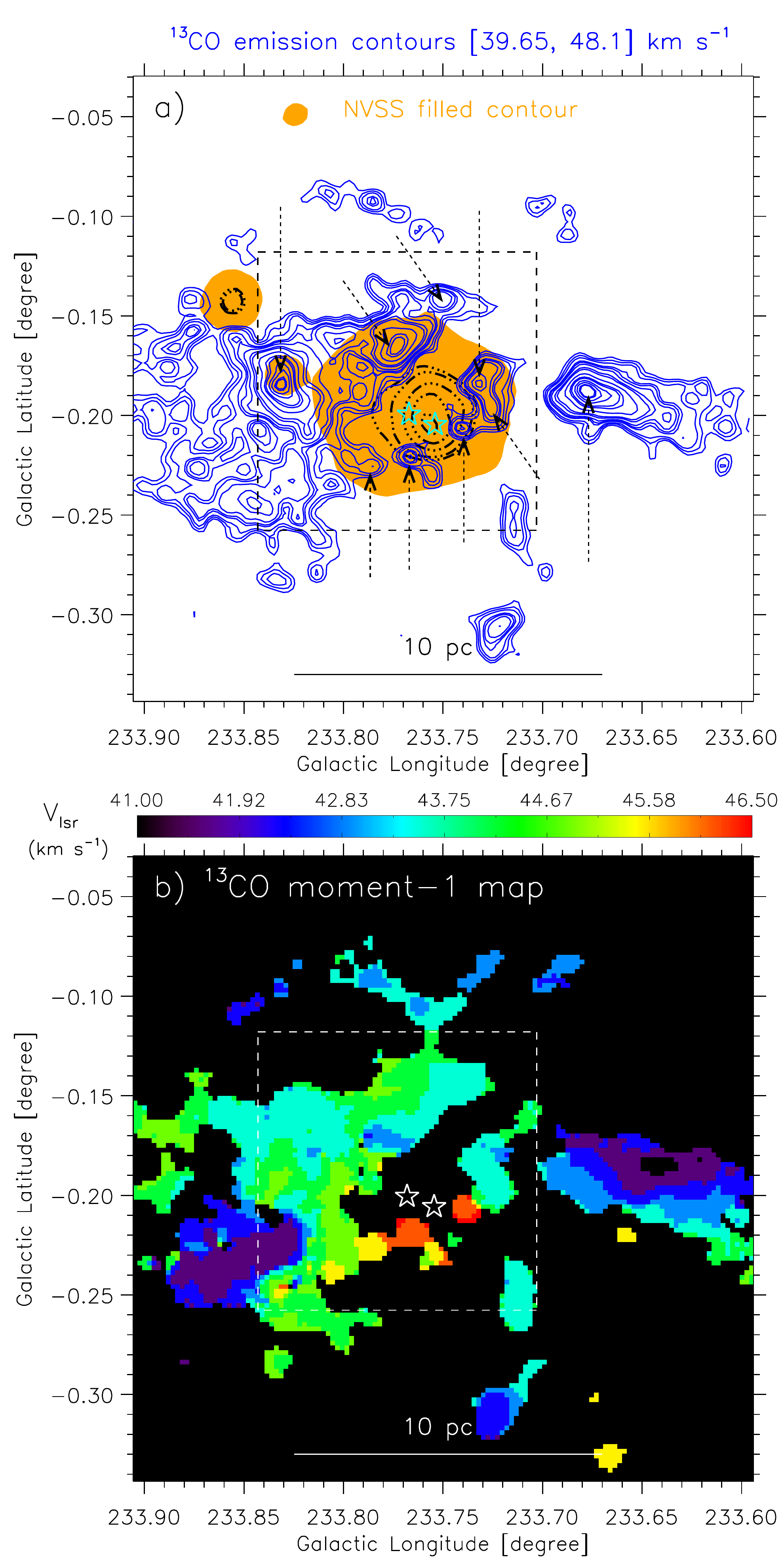}
\caption{a) Overlay of the FUGIN $^{13}$CO(J=1$-$0) emission on the filled NVSS contour map (at 2.3 mJy beam$^{-1}$; in orange). The $^{13}$CO integrated velocity range is from 
39.65 to 48.1 km s$^{-1}$ (1$\sigma$ $\sim$0.85 K km s$^{-1}$). The $^{13}$CO contours (in blue) are 
3.6, 4.3, 5, 6.2, 6.7, 7.5, 9, 10, 13, 16, 17.3, 20, 22, 24, and 24.6 K km s$^{-1}$. 
We also show in broken-line black contours the 50, 60, 85, 110, and 135 mJy beam$^{-1}$ flux levels at 1.4 GHz. Several molecular condensations surrounding massive stars are highlighted by black arrows (see also Figure~\ref{ufig1}b). 
b) Intensity-weighted mean velocity (moment-1) map of $^{13}$CO.
In each panel, stars show the positions of two massive O-type stars, and the scale bar referring to 10 pc (at a distance of 3.7 kpc) is displayed.} 
\label{ufig2}
\end{figure*}

\begin{figure*}
\epsscale{1.1}
\plotone{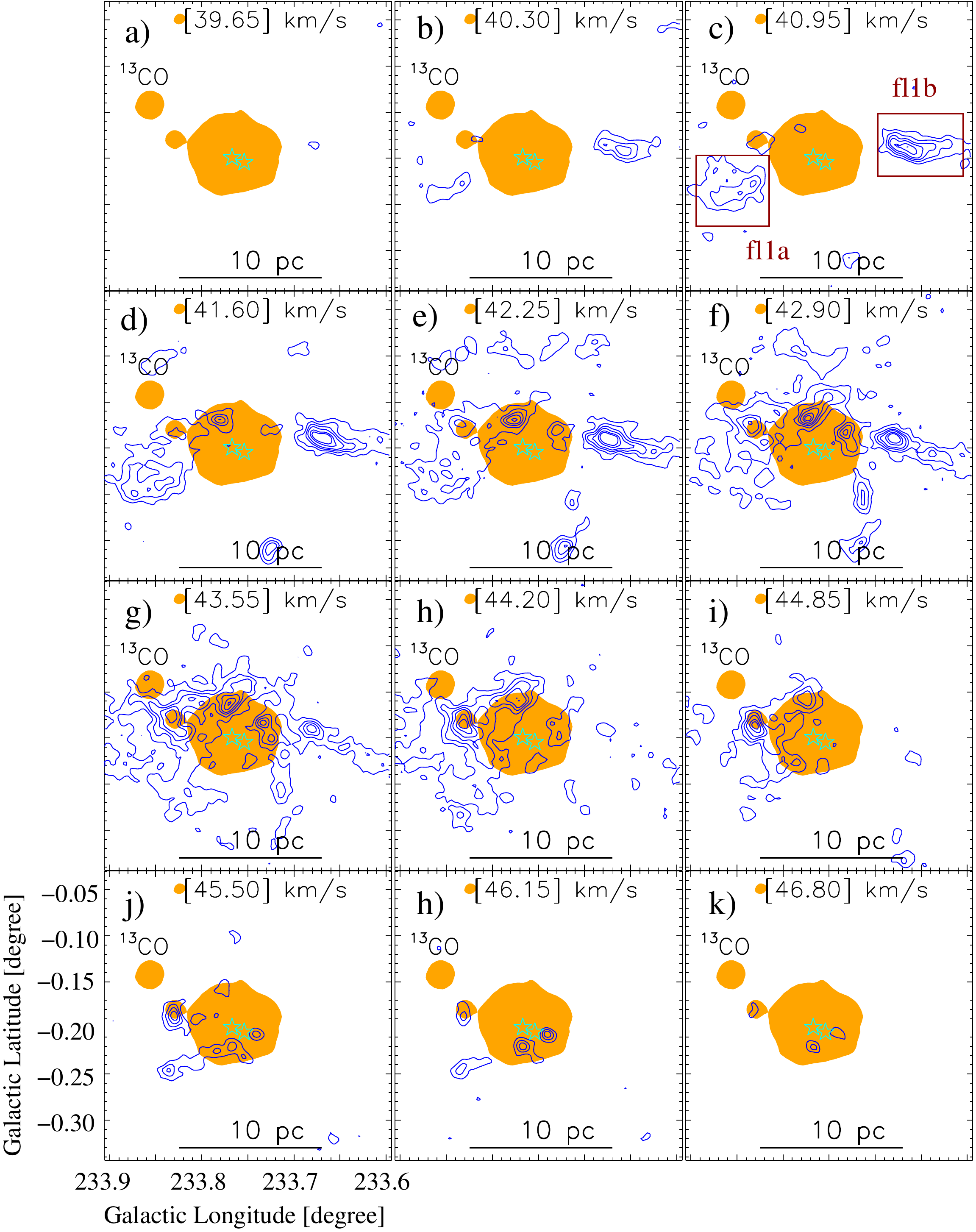}
\caption{Velocity channel contours of $^{13}$CO emission. 
The velocity of molecular emission is marked in each panel (in km s$^{-1}$). 
The contour levels of $^{13}$CO are presented with the levels of 1, 2, 3, 4, 5.5, and 6.5 K km s$^{-1}$. 
In panel ``c",  solid boxes highlight both ends of the filament (i.e., fl1a and fl1b) traced in the {\it Herschel} column density map. 
In each panel, the NVSS filled contour is shown in the background with a level of 2.3 mJy beam$^{-1}$ (see also Figure~\ref{ufig2}a). 
In all panels, stars show the positions of two massive O-type stars.} 
\label{fig7}
\end{figure*}

\begin{figure*}
\epsscale{1}
\plotone{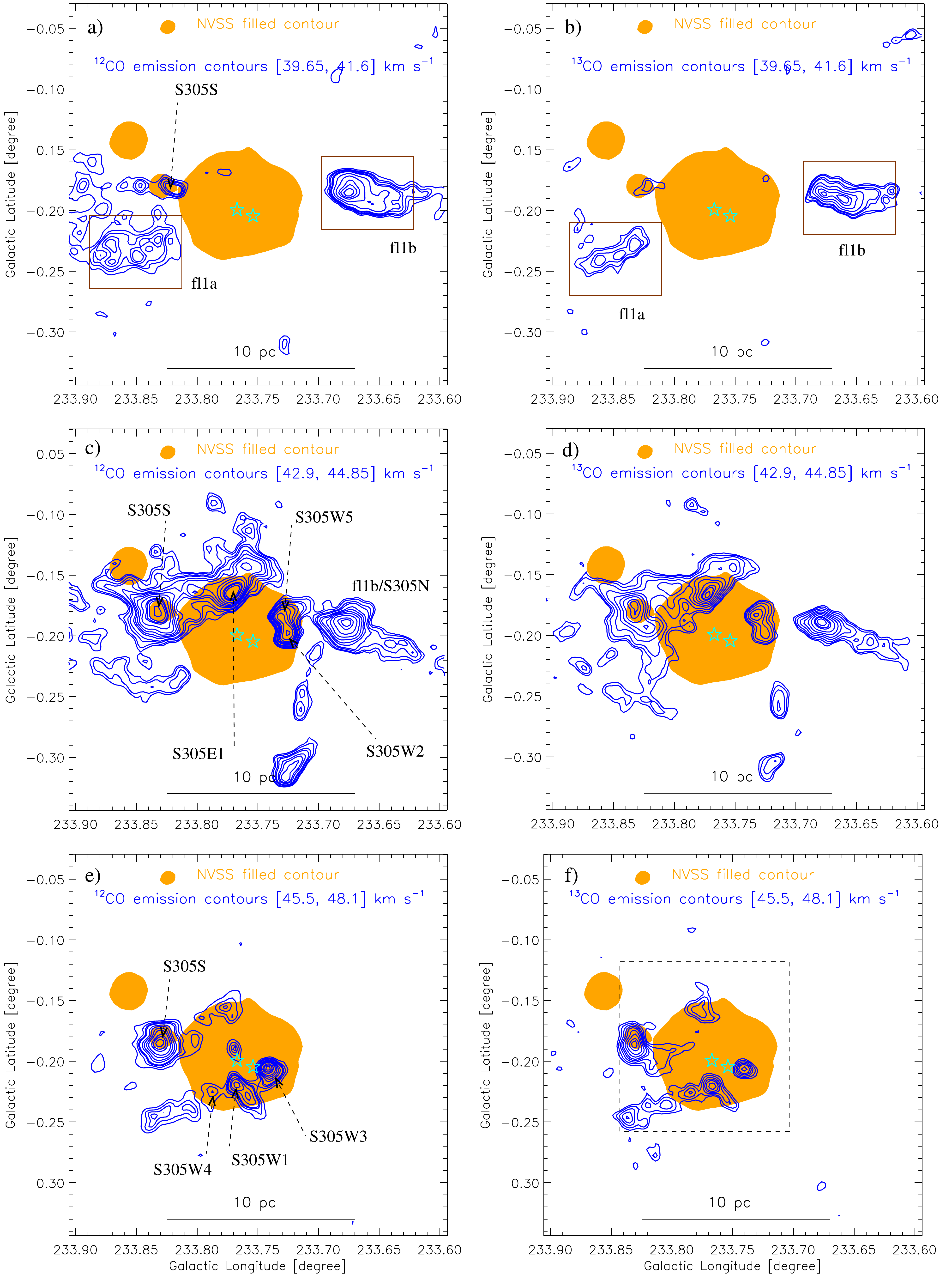}
\caption{\textbf{Left panels (i.e., a, c, e):} Velocity channel contours of $^{12}$CO emission at different velocity intervals.
In panel ``a", the contours of $^{12}$CO are 5.5, 7, 8, 9.5, 10.5, 14, 18, 22, 24, and 26 K km s$^{-1}$. 
In panel ``c", the contours of $^{12}$CO are 13.8, 15, 17, 18.5, 21.5, 23.5, 26, 29, 33, 38, 40, 43, and 
49 K km s$^{-1}$. 
In panel ``e", the contours of $^{12}$CO are 8.5, 11, 13, 14.2, 18, 21, 25, 29, 36, 40, and 45 
K km s$^{-1}$. 
\textbf{Right panels (i.e., b, d, f):} Velocity channel contours of $^{13}$CO emission at different velocity intervals.
In panel ``b", the contours of $^{13}$CO are 1.6, 1.95, 2.6, 3, 4, 5, 6, and 6.6 K km s$^{-1}$. 
In panel ``d", the contours of $^{13}$CO are 3, 3.5, 4.3, 5.5, 6.8, 8.3, 10, 12, 12.8, and 14 K km s$^{-1}$. 
In panel ``f", the contours of $^{13}$CO are 1.8, 2.4, 3.2, 4, 4.8, 6, 7.3, 8.8, and 9.8 K km s$^{-1}$. 
In panel ``f", a dashed box (in black) encompasses the area shown in Figures~\ref{fig4}a,~\ref{fig4}b, and~\ref{fig4}c.
In each panel, the NVSS filled contour is shown in the background with a level of 2.3 mJy beam$^{-1}$ (see also Figure~\ref{ufig2}a). 
In panels ``a" and ``b", solid boxes represent both ends of the 
filament (i.e., fl1a and fl1b). 
In all panels, stars show the positions of two massive O-type stars. The velocity interval of molecular emission (in km s$^{-1}$) is indicated in each panel.} 
\label{hfig7}
\end{figure*}

In Figure~\ref{ufig2}a, we present the contours of $^{13}$CO intensity map at [39.65, 48.1] km s$^{-1}$, which are overlaid on the NVSS radio continuum contours (in black). In the background, we also show the NVSS filled contour (in orange) at 2.3 mJy beam$^{-1}$. 
Molecular condensations are also observed in the $^{13}$CO map (see arrows in Figure~\ref{ufig2}a), and are similar as seen in the $^{12}$CO map (see also Figure~\ref{ufig1}b). 
Figure~\ref{ufig2}b displays the moment-1 map of $^{13}$CO, allowing us to study the intensity-weighted mean 
velocity of the emitting gas. Hence, Figures~\ref{ufig2}a and~\ref{ufig2}b enable us to 
compare the moment-0 map against the moment-1 map of $^{13}$CO. 
Radial velocity information toward the filament, the IR shell, and the horseshoe envelope can be inferred in Figure~\ref{ufig2}b, which seem to be depicted in different velocity ranges of molecular gas. 
To further explore this observational finding, Figure~\ref{fig7} displays the velocity channel maps of $^{13}$CO (blue contours) from 39.65 to 46.8 km s$^{-1}$. These maps also show that the molecular gas is found toward the observed features at different velocity intervals. 
In addition, in Figure~\ref{hfig7} we show the integrated molecular maps of $^{12}$CO (see left panels) and $^{13}$CO (see right panels) at different velocity intervals. These velocity intervals are [39.65, 41.6], [42.9, 44.85], and [45.5, 48.1] km s$^{-1}$. 
In each panel of Figures~\ref{fig7} and~\ref{hfig7}, the distribution of molecular gas is presented against the NVSS filled contour (in orange). In the maps of $^{12}$CO and $^{13}$CO at [39.65, 41.6] km s$^{-1}$, both ends of the filament (i.e. fl1a and fl1b) are traced (see Figures~\ref{hfig7}a and~\ref{hfig7}b). 
The molecular gas at [42.9, 44.85] km s$^{-1}$ is found toward the IR shell (see Figures~\ref{hfig7}c and~\ref{hfig7}d). 
Regularly spaced molecular condensations are also seen in the direction of the edges of the IR shell. 
The molecular gas toward the horseshoe envelope is depicted at [45.5, 48.1] km s$^{-1}$ (see Figures~\ref{hfig7}e and~\ref{hfig7}f). Assuming a global expansion model, the filament appears to be located in the front of both the IR shell and the horseshoe envelope. 
Furthermore, the edges of the IR shell seem to be located in the front of the horseshoe envelope. One may note that 
the clump ``S305S" is traced in all velocity intervals. 
Considering the distribution of $^{12}$CO and $^{13}$CO at [42.9, 44.85] and [45.5, 48.1] km s$^{-1}$, a signature of an expanding shell of molecular gas is evident in the S305 H\,{\sc ii} region (see panels ``c", ``e" and/or panels ``d", ``f"). Detailed discussions on all these observed findings are presented in Section~\ref{sec:disc}.
\subsection{High resolution radio continuum maps}
\label{subsec:radio}
\subsubsection{Extended radio morphology of S305}
\label{subsec:radio1}
\begin{figure*}
\epsscale{0.5}
\plotone{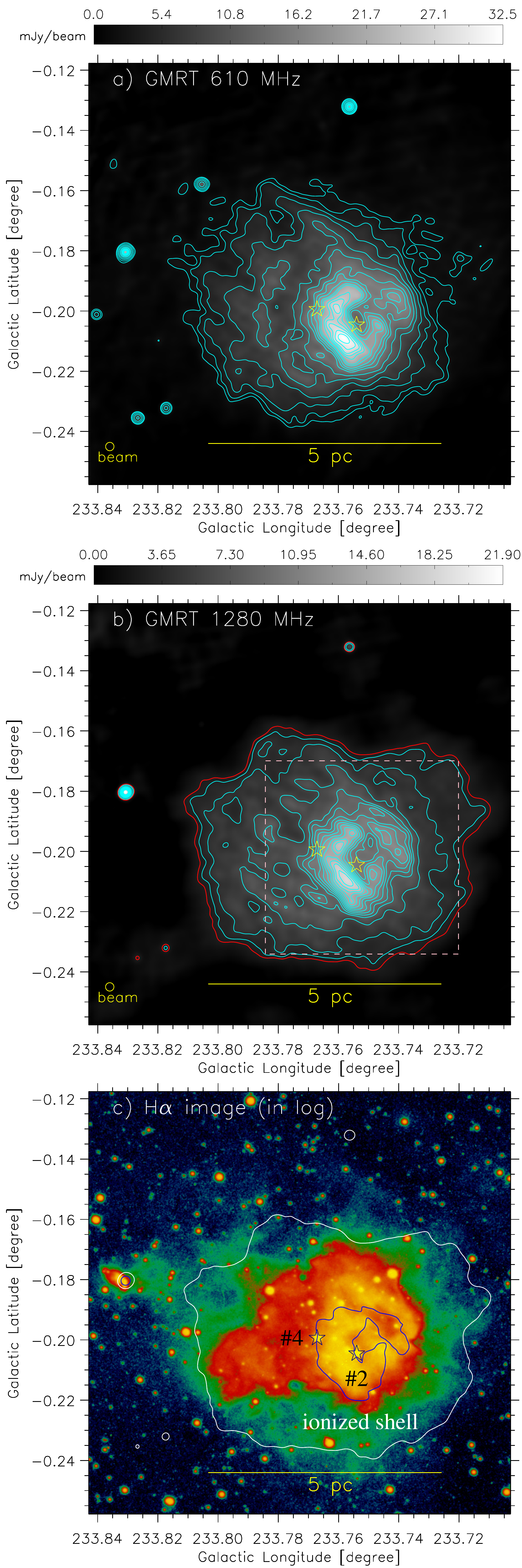}
\caption{a) Overlay of the GMRT 610~MHz continuum contours on 
the 610 MHz gray scale map (resolution $\sim$ 10$''$ $\times$ 10$''$; see dashed box in Figure~\ref{hfig7}f). 
The GMRT 610 MHz continuum contour levels (in cyan) are 2.2, 3, 4.5, 6.5, 9.5, 12, 15, 16.5, 18, 
20, 21.3, 24, 25.5, 27, 29, and 31 mJy beam$^{-1}$ (with $\sigma \sim0.46$ mJy beam$^{-1}$). b) Overlay of the GMRT 1280 MHz continuum
contours on the 1280 MHz gray scale map (resolution $\sim$ 10$''$ $\times$ 10$''$). 
The GMRT 1280 MHz continuum contour levels (in cyan) are 3, 4, 6, 8, 9, 10, 11, 12, 13, 14, 15, 16, 17, 
and 18 mJy beam$^{-1}$ (with $\sigma$ $\sim$0.74 mJy beam$^{-1}$). 
The GMRT 1280 MHz continuum level at 2.5 mJy beam$^{-1}$ is shown with a red contour. 
A pink dashed box encompasses the area shown in Figures~\ref{fig5}a,~\ref{fig5}b,~\ref{fig5}c, 
and~\ref{fig5}d. 
c) Overlay of the GMRT 1280~MHz continuum contours (at 2.5 and 10.5 mJy beam$^{-1}$) on the H$\alpha$ image. 
In all panels, stars show the positions of two massive O-type stars (see also Figure~\ref{ufig1}a).}
\label{fig4}
\end{figure*}

The H$\alpha$ image and the NVSS radio continuum emission enable us to study the distribution of the ionized gas in S305. 
Due to a coarse beam size, the NVSS radio continuum map does not allow to compare the inner ionized feature traced in the optical 
H$\alpha$ image. Hence, in order to explore the inner morphology of the ionized emission in S305, we utilize the GMRT radio continuum maps. Figure~\ref{fig4}a shows the GMRT 610 MHz continuum map (beam size $\sim$10$''$ $\times$ 10$''$) overlaid 
with the 610 MHz continuum contours. 
In Figure~\ref{fig4}b, we display the GMRT 1280 MHz continuum map (beam size $\sim$10$''$ $\times$ 10$''$) superimposed with the 1280 
MHz continuum contours. In Figures~\ref{fig4}a and~\ref{fig4}b, we can compare the observed radio morphology of S305 in two GMRT radio bands. Both GMRT maps display a similar radio structure, which is an extended sphere-like morphology (extension $\sim$5.5 pc). Based on a visual inspection, we find the brighter region between the positions of the two O stars in both the radio maps, suggesting the existence of a compact radio clump. The extended sphere-like radio morphology is found well within the IR shell. 
Furthermore, both GMRT radio continuum maps are compared with the morphology traced in the H$\alpha$ image (see Figure~\ref{fig4}c). In Figure~\ref{fig4}c, the GMRT 1280 MHz continuum contours at 2.5 and 10.5 mJy beam$^{-1}$ are also overlaid on the H$\alpha$ image.
The GMRT 1280 MHz continuum contour at 2.5 mJy beam$^{-1}$ shows the extension of the S305 H\,{\sc ii} region, while the GMRT contour at 10.5 mJy beam$^{-1}$ confirms 
the existence of the ionized shell toward the center of the horseshoe envelope containing the massive O-type stars.

Using the radio continuum map at 1280 MHz, the integrated flux density ($S_{\nu}$) and the radius ($R_\mathrm{HII}$) 
of the H\,{\sc ii} region are estimated to be $\sim$4.0 Jy and $\sim$2.75 pc, respectively. 
The {\it clumpfind} IDL program \citep{williams94} is employed to obtain the value of $S_{\nu}$, which is estimated 
within the contour of 2.5 mJy beam$^{-1}$ at 1280 MHz (see a red contour in Figure~\ref{fig4}b). 
Using the radio continuum map at 610 MHz, the values of $S_{\nu}$ and $R_\mathrm{HII}$ are also computed to be $\sim$5.0 Jy and 2.65 pc, respectively (see Figure~\ref{fig4}a).
 
In the S305 H\,{\sc ii} region, it seems that the distribution of the ionized emission is not uniform, and 
most of the ionized emission is traced mainly toward the positions of two massive O-type 
stars (see Figures~\ref{fig4}a,~\ref{fig4}b, and~\ref{fig4}c), where the ionized shell is investigated in the H$\alpha$ image.  
\subsubsection{Spectral index map}
\label{subsec:alpha}
\begin{figure*}
\epsscale{1.15}
\plotone{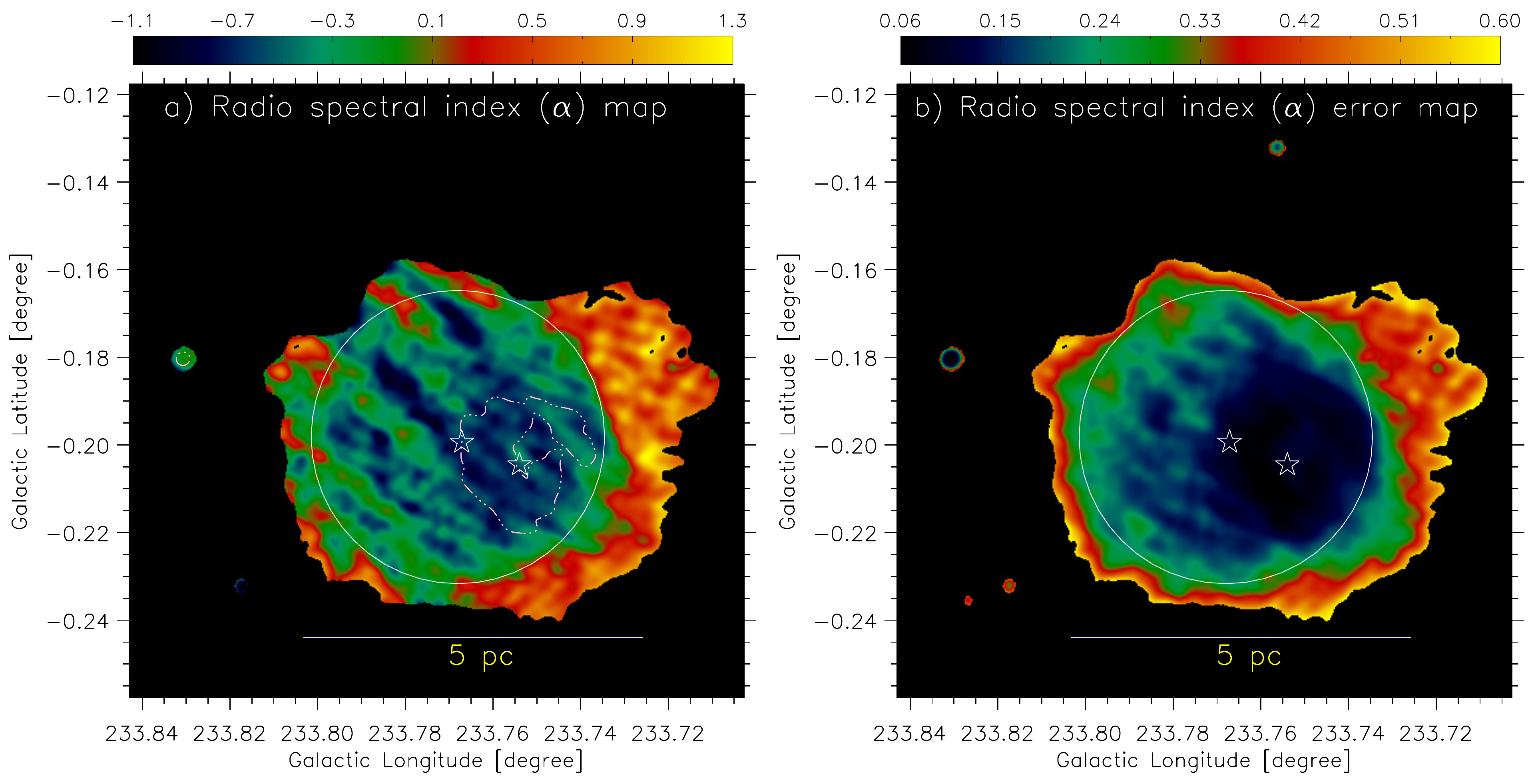}
\caption{a) GMRT spectral index map (resolution $\sim$10$''$) of S305 generated for pixels above 3$\sigma$ level in the 610 and 1280 MHz bands. The GMRT 1280~MHz continuum emission is also shown by a dotted-dashed contour at 10.5 mJy beam$^{-1}$. 
Regions with non-thermal $\alpha$ are greenish-blue, while regions consistent with thermal emission are red-yellow. 
b) GMRT spectral index error map of S305. In both panels, stars show the positions of two 
massive O-type stars (see also Figure~\ref{ufig1}a), and a big circle (in white) highlights the central region of S305.}
\label{fig6}
\end{figure*}

We have computed a spectral index ($\alpha$) map of the S305 H\,{\sc ii} region using the GMRT 
radio maps at 610 and 1280 MHz. The analysis of the radio spectral index 
is performed using only these two data points.  
The spectral index, defined as $F_\mathrm{\nu}$ $\propto$ $\nu^{\alpha}$, 
enables us to examine the nature (thermal and non-thermal) of the radio continuum emission in S305. 
Here, $\nu$ is the frequency of observation, and $F_\mathrm{\nu}$ is the corresponding observed flux density. Figures~\ref{fig6}a and~\ref{fig6}b present the radio spectral index map 
and its corresponding error map, respectively. The spectral index map is 
constructed by estimating $\alpha$ in each pixel using the AIPS software. 
For this purpose, both GMRT maps were produced using the visibilities in the 
same \textit{uv} range of 
0.088--52.24 k$\lambda$, the same pixel size of 1$''$, and the same angular resolution of 10$''$. 
In both maps, we considered only pixels with a high signal-to-noise (above 3$\sigma$).
The error map shows the uncertainties in $\alpha$ for each pixel.  

The spectral index value in S305 varies between $-1.1$ to 1.3, with errors typically 
$\lesssim 0.2$ and of $\sim$0.5 at most. 
GMRT maps also show that the central region of S305 is dominated by the ionized emission, and is highlighted by a 
big circle in Figures~\ref{fig6}a and~\ref{fig6}b. 
The average spectral index within this region is $-0.45$.  
The spectral index map is also overlaid with the GMRT 1280 MHz radio continuum contour (at 10.5 mJy beam$^{-1}$). We discuss the interpretation of these findings in Section~\ref{sec:disc_nonthermal}. 

Considering the observed spectral index range (i.e., [$-1.1$, +1.3]) in the S305 H\,{\sc ii} region, the observed average spectral index of $-0.45$ is consistent with a combination of thermal and non-thermal emission. As the non-thermal emission has a negative spectral index, it is significantly higher at low radio frequencies, whereas the thermal emission either has a weak dependence on frequency (if the emitter is optically thin) or it is stronger at higher frequencies (if the emitter is optically thick). Unfortunately, disentangling both emission components is problematic due to the limited frequency coverage of the radio data. We therefore adopt a simplified approach and then discuss possible caveats in this.

We assume that there are two emission components, one thermal from the ionized gas and other one non-thermal from the relativistic electrons, such that $F_\mathrm{obs}(\nu) = F_\mathrm{th}(\nu) + F_\mathrm{NT}(\nu)$. The gas in the H\,{\sc ii} region is most likely optically thin so that $F_\mathrm{th}(\nu) \propto \nu^{-0.1}$. For the non-thermal synchrotron emission we consider a spectral index $\alpha_\mathrm{NT} = -0.8$, which is about the most negative value observed in the spectral index map (Figure~\ref{fig6}a), so that $F_\mathrm{NT}(\nu) \propto \nu^{-0.8}$. By doing this, we can calculate the normalization constants for both emission components from the observed fluxes at 610 and 1280 MHz (5~Jy and 4~Jy, respectively). We obtain $F_\mathrm{NT}(1280~\mathrm{MHz}) = 0.95$~Jy and $F_\mathrm{th}(1280~\mathrm{MHz}) = 3.05$~Jy. Thus, 76\% of the observed flux at 1280 MHz comes from the ionized gas, whereas at 610 MHz it contributes with 66\% of the observed flux. These numbers can vary slightly if one assumes different values of $\alpha_\mathrm{NT}$: for $\alpha_\mathrm{NT}=-1$ up to 82\% of the observed flux at 1280~MHz is thermal, whereas for $\alpha_\mathrm{NT}=-0.6$ this fraction drops to 65\%.
\subsubsection{Lyman continuum photons}
\label{subsec:zzradio1}
One can perform an exercise to determine the number of Lyman continuum photons ($N_\mathrm{UV}$) using the radio continuum map, which enables us to 
infer the spectral type of the powering candidate responsible for the observed radio emission. In this connection, one can use the 
following equation \citep{matsakis76} to compute the value of $N_\mathrm{UV}$:
\begin{equation}
\begin{split}
N_\mathrm{UV} (s^{-1}) = 7.5\, \times\, 10^{46}\, \left(\frac{F_\mathrm{th}(\nu)}{\mathrm{Jy}}\right)\left(\frac{D}{\mathrm{kpc}}\right)^{2} 
\left(\frac{T_{e}}{10^{4}\mathrm{K}}\right)^{-0.45} \\ \times\,\left(\frac{\nu}{\mathrm{GHz}}\right)^{0.1}
\end{split}
\end{equation}
\noindent where $F_\mathrm{th}(\nu)$ is the estimated thermal flux density in Jy, $D$ is the distance in kpc, 
$T_{e}$ is the electron temperature, and $\nu$ is the frequency in GHz. 
Based on our analysis of the radio spectral index map, we obtain the value of $F_\mathrm{th}(\nu)$ = 3.05~Jy at 1280 MHz. 
In the calculation of $N_\mathrm{UV}$, one assumes that all the ionizing flux is generated by only a single massive OB star. 
Adopting the values of $T_{e}$ = 10\,000~K and D = 3.7 kpc in equation~1, 
we estimate $N_\mathrm{UV}$ (or $\log{N_\mathrm{UV}}$) to be $\sim$3.2 $\times$ 10$^{48}$ s$^{-1}$ (48.51) for the S305 H\,{\sc ii} region. 

Based on the multi-wavelength images, the S305 H\,{\sc ii} region can be considered as a dusty H\,{\sc ii} region. 
As mentioned earlier, there are two massive O8.5V (\#4) and O9.5V (\#2) stars known in S305 \citep[e.g.,][]{pandey20}. 
In Figure~\ref{fig4}c, the O8.5V star (\#4) appears to be located geometrically near the center of the S305 H\,{\sc ii} region, while the O9.5 star (\#2) is seen toward the center of the horseshoe envelope or the ionized shell. Following the theoretical work of \citet{panagia73}, the values of $N_\mathrm{UV}$ (or $\log{N_\mathrm{UV}}$) are reported to be 
$\sim$2.8 $\times$ 10$^{48}$ s$^{-1}$ (48.45) and $\sim$1.2 $\times$ 10$^{48}$ s$^{-1}$ (48.08) for O8.5V and O9.5V stars, respectively.
If we add these two values of $N_\mathrm{UV}$ concerning O8.5V and O9.5V stars, then the combined value of $N_\mathrm{UV}$ (or $\log{N_\mathrm{UV}}$)
is equivalent to $\sim$4.02 $\times$ 10$^{48}$ s$^{-1}$ (48.6). 
This value is higher than the derived value of $N_\mathrm{UV}$ of the S305 H\,{\sc ii} region. 
We suggest that the remaining value of $N_\mathrm{UV}$ (i.e., 4.02 $\times$ 10$^{48}$ $-$ 3.2 $\times$ 10$^{48}$ $\sim$8.2 $\times$ 10$^{47}$ s$^{-1}$) could be absorbed by dust grains prior to contributing to the ionization \citep{inoue01,inoue01a,binder18}. In Figure~\ref{ufig1}a, several dust pillars apparent on the rim of the infrared shell 
appear oriented in the direction of star \#4, but not star \#2, suggesting that the star \#4 is indeed the dominant source of feedback in the region and is responsible for ionizing the diffuse component of the S305 H\,{\sc ii} region. However, the horseshoe envelope or the ionized shell appears to be ionized primarily by the star \#2 (see Figure~\ref{lfig1}). 

Using the values of $N_\mathrm{UV}$ and $R_\mathrm{HII}$, we also computed the dynamical age ($t_\mathrm{dyn}$) of the entire S305 H\,{\sc ii} region using the following equation \citep{dyson80}: 
\begin{equation}
t_\mathrm{dyn} = \left(\frac{4\,R_{s}}{7\,c_{s}}\right) \,\left[\left(\frac{R_\mathrm{HII}}{R_{s}}\right)^{7/4}- 1\right] 
\end{equation}
where $c_{s}$ is the isothermal sound velocity in the ionized gas ($c_{s}$ = 11 km s$^{-1}$; \citet{bisbas09}), 
$R_\mathrm{HII}$ is previously defined, and $R_{s}$ is the radius of the Str\"{o}mgren sphere (= (3 N$_\mathrm{UV}$/4$\pi n^2_{\rm{0}} \alpha_{B}$)$^{1/3}$, where 
the radiative recombination coefficient $\alpha_{B}$ =  2.6 $\times$ 10$^{-13}$ (10$^{4}$ K/T)$^{0.7}$ cm$^{3}$ s$^{-1}$ \citep{kwan97}, 
N$_\mathrm{UV}$ is defined earlier, and ``$n_\mathrm{0}$'' is the initial particle number density of the ambient neutral gas). 
The analysis assumes that the H\,{\sc ii} regions are uniform and spherically symmetric. 
Taking into account the typical value of $n_\mathrm{0}$ (= 10$^{3}$--10$^{4}$ cm$^{-3}$), the dynamical age of the S305 H\,{\sc ii} region is calculated to be $\sim$0.5--1.7 Myr.
\subsubsection{Inner radio morphology of S305: ionized shell}
\label{subsec:radio1}
\begin{figure*}
\epsscale{0.9}
\plotone{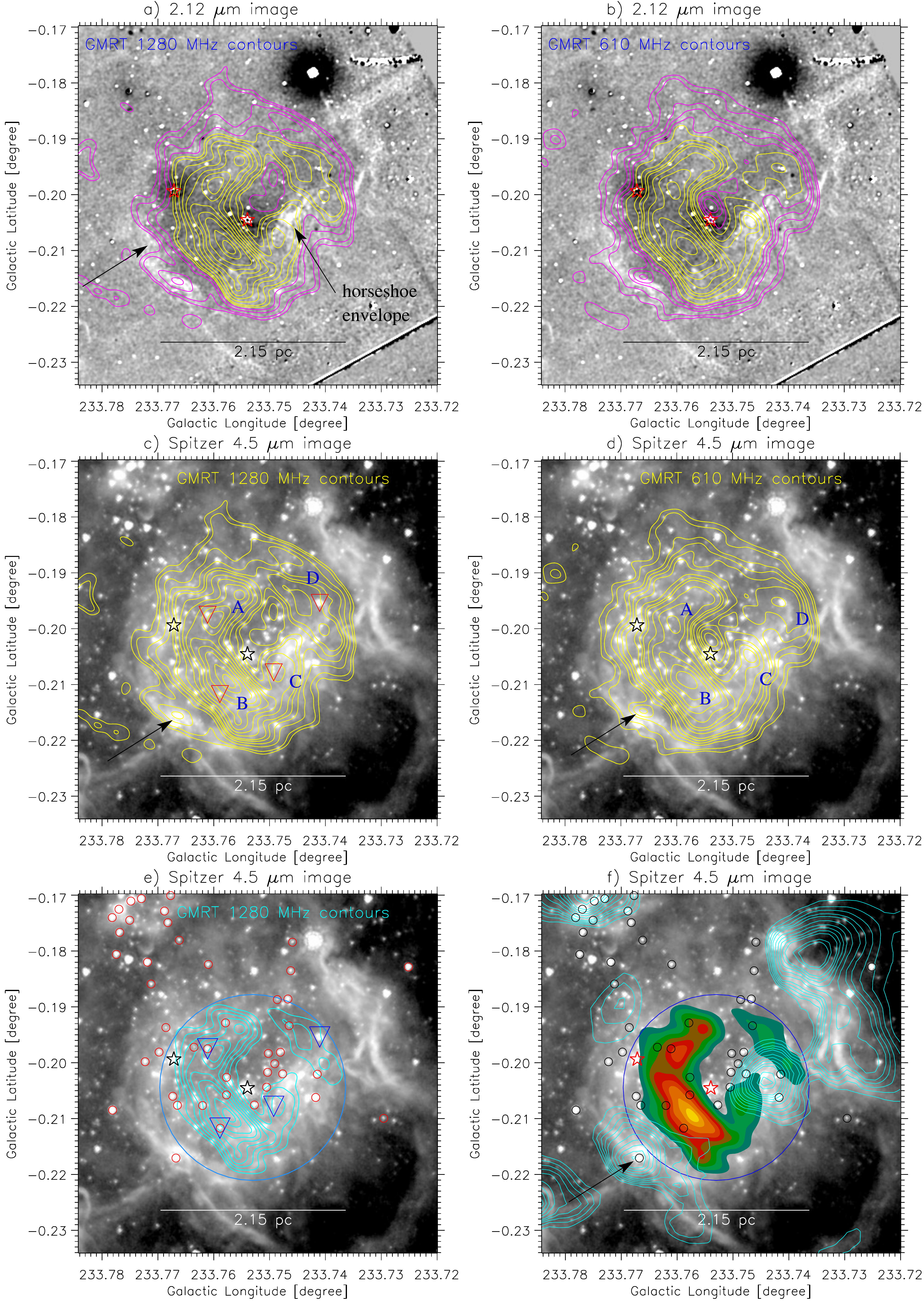}
\caption{a) Overlay of the GMRT 1280~MHz continuum contours (in magenta and yellow) on the continuum subtracted H$_{2}$ image 
at 2.12 $\mu$m (see a dashed box in Figure~\ref{fig4}b and also Figure~\ref{vfg1}b). 
The contour levels in magenta are 7.5, 8, 9, and 9.8 mJy beam$^{-1}$, while the yellow contours are 10.5, 11, 12, 13, 14, 15, 16, 17, 18, and 18.5 mJy beam$^{-1}$. 
b) Overlay of the GMRT 610 MHz continuum contours (in magenta and yellow) on the continuum subtracted H$_{2}$ image at 2.12 $\mu$m. 
The magenta contour levels are 10.5, 11.5, 13.5, 14.5, and 15.5 mJy beam$^{-1}$, while the yellow ones are 16.5, 18, 20.5, 23, 25, 26, 29, and 31 mJy beam$^{-1}$. 
c) Overlay of the GMRT 1280 MHz continuum contours (in yellow) on the {\it Spitzer} 4.5 $\mu$m image (see Figure~\ref{fig5}a). d) Overlay of the GMRT 610 MHz continuum contours (in yellow) on the {\it Spitzer} 4.5 $\mu$m image (see Figure~\ref{fig5}b). 
e) The panel is the same as in Figure~\ref{fig5}c, but it is overlaid with the positions of YSOs \citep[see small red circles; from][]{pandey20}. The GMRT 1280 MHz continuum contours (in cyan) are shown with the levels 
of 11, 12, 13, 14, 15, 16, 17, 18, and 18.5 mJy beam$^{-1}$.
f) The panel is the same as in Figure~\ref{fig5}e, but it is overlaid the $N(\mathrm H_2)$ contours (in cyan).
The $N(\mathrm H_2)$ contours (in cyan) are presented with the levels of (7.0, 8.0, 9.0, 10.0, 11.0, 13.0, 15.0, 17.0, 19.0, 21.0, 26.0, and 28.5) $\times$ 10$^{21}$ cm$^{-2}$ (see also Figure~\ref{fig2}b). 
The GMRT 1280 MHz continuum emission is displayed by filled contours. 
In panels ``c" and ``d", four selected radio peaks are labeled (``A--D"). 
In panels ``c" and ``e", the position of a point-like source toward each radio peak is shown by an upside down triangle. In the last two panels, a big circle (extension $\sim$2.15 pc) highlights a ring-like radio morphology around a massive O9.5V (\#2) star. In all panels, stars show the positions of two massive O-type stars (see also Figure~\ref{ufig1}a).} 
\label{fig5}
\end{figure*}

In order to further examine the ionized shell, in Figure~\ref{fig5}, a zoomed-in view of the area highlighted by 
a broken box in Figure~\ref{fig4}b is shown. 
Figures~\ref{fig5}a and~\ref{fig5}b show the overlay of the GMRT 1280 and 610 MHz continuum emission (see magenta and yellow contours) 
on the continuum-subtracted H$_{2}$ image at 2.12 $\mu$m, respectively.
Figures~\ref{fig5}c and~\ref{fig5}d show the overlay of the GMRT 1280 and 610 MHz continuum emission (see yellow contours) 
on the {\it Spitzer} 4.5 $\mu$m image, respectively. 
In Section~\ref{sec:morph1}, using the continuum-subtracted H$_{2}$ map and the {\it Spitzer} 4.5 $\mu$m image, we have discussed about the existence of the horseshoe envelope feature in the direction of two massive O-type stars. Both radio emission contours show a complete shell-like structure, which appears to follow the horseshoe envelope (see magenta contours in Figures~\ref{fig5}a and~\ref{fig5}b, and see also 
Figures~\ref{fig5}c and~\ref{fig5}d). Additionally, in the direction of the center of the horseshoe envelope (i.e., gas deficient cavity) 
or the ionized shell, the GMRT radio maps display a curved partial-shell structure containing radio peaks (see yellow contours in Figures~\ref{fig5}a and~\ref{fig5}b, and also Figures~\ref{fig5}c and~\ref{fig5}d). Based on a visual inspection, at least six radio peaks are seen in the GMRT 1280 MHz continuum map, while one can find 
four radio peaks in the map at 610 MHz.  
Hence, we select four radio peaks/ionized clumps (i.e., ``A--D"), which are observed in both the GMRT maps (see labels in Figures~\ref{fig5}c and~\ref{fig5}d). 

In general, it may be possible that these ionized clumps could be local overdensities in the ionized/shocked interstellar medium (ISM), where new stars are not yet born. 
In this connection, we examined the {\it Spitzer} images to find out a point source or a compact nebula in the direction of each ionized clump. 
Based on the availability of photometric data of point-like sources, at least one point source is identified toward each ionized clump (see upside down triangles in Figures~\ref{fig5}c and~\ref{fig5}e). 
The analysis of the color-magnitude space of these sources shows the presence of a B1.5-B2 type star associated with clump ``D".
However, no massive star candidate is found toward other three ionized clumps (i.e., ``A--C").
In Figure~\ref{fig5}e, we overlay the positions of YSOs and the GMRT 1280 MHz continuum emission contours on 
the {\it Spitzer} 4.5 $\mu$m image. The selected point-like sources toward the ionized clumps ``A" and ``B" coincide with the positions of YSOs. 
Hence, these YSOs are likely to be low-mass star candidates. In Figure~\ref{fig5}f, we also present the {\it Herschel} column density contours, the positions of YSOs (see small circles), and the GMRT filled contours at 1280 MHz on 
the {\it Spitzer} 4.5 $\mu$m image. There is no YSO seen toward the ionized clump ``C", 
which also appears away from the {\it Herschel} clump. 

The horseshoe envelope is associated with the {\it Herschel} clumps and the radio continuum peaks/clumps (see clumps ``C" and ``D", and also arrows in Figures~\ref{fig5}c, ~\ref{fig5}d, and~\ref{fig5}f). A big circle (extension $\sim$2.15 pc) is also drawn in Figures~\ref{fig5}e and~\ref{fig5}f, where the average spectral index is determined to be about $-0.6$ (see Figure~\ref{fig6}a). 

Together, it is evident that there are overdensities of the ionized emission distributed in the direction of two massive O-type stars, which are surrounded by the horseshoe envelope.
\section{Discussion}
\label{sec:disc}
%
\subsection{Star formation scenario}
\label{sec:disc}
In our selected target site S305, IR observations from the {\it Spitzer} and {\it Herschel} facilities 
have revealed the extended sphere-like IR shell, the elongated filament, and the horseshoe envelope.
Based on the observed radial velocities, the moment-1 map of $^{13}$CO clearly distinguishes these three different observed 
structures in S305 (see Section~\ref{sec:gas}). It is worth noting that the positions of these features are shifted with 
respect to each other along the line of sight. 
The radio continuum emission is well distributed within the extended IR shell, and shows a complete shell-like structure. 
Our analysis of the radio continuum maps shows that the S305 H\,{\sc ii} region is excited by 
two O-type massive stars (see Section~\ref{subsec:radio1}). However, the observed ionized emission is mainly dominated 
toward the center of the horseshoe envelope. 
The {\it Herschel} temperature map reveals the existence of the warm dust emission at $T_\mathrm{d}$ = 17.5--27~K 
toward the IR shell. This particular result suggests the impact/heating by two massive 
O-type stars to their surroundings (see Section~\ref{sec:dmorph1}). 
It is supported with the fact that the ionizing/far-UV radiation from massive O-type stars is the major heating agent in S305. It illustrates an interaction between the molecular environment and the S305 H\,{\sc ii} region. 
Most recently, \citet{pandey20} estimated three pressure components (i.e., pressure of an H\,{\sc ii} 
region $(P_{\rm HII})$, radiation pressure (P$_{\rm rad}$), and stellar wind ram pressure (P$_{\rm wind}$)) driven 
by two massive O-type stars in S305, which allowed them to study their feedback. 

Several molecular condensations, dust clumps, and YSOs are found in the direction of the edges of the IR shell 
associated with the S305 H\,{\sc ii} region (see Section~\ref{sec:morph}). 
In Figure~\ref{fig2}a, we find that at least five dust clumps (mass range $\sim$270--1555 M$_{\odot}$; 
density range $\sim$1880--3790 cm$^{-3}$) seem to be nearly regularly spaced along the sphere-like 
shell surrounding the ionized emission. The average volume density of each clump is also computed 
using the equation $n_{\mathrm H_2} = 3M_\mathrm{clump}/(4\pi R_\mathrm{clump}^3 \mu_{\rm H_2} m_{\rm H})$, assuming 
that each clump has a spherical geometry. Here, $m_{\rm H}$ is the mass of an hydrogen atom and the 
mean molecular weight $\mu_{\rm H_2}$ is assumed to be 2.8. 
The analysis of the FUGIN molecular line data hints an expansion of the S305 H\,{\sc ii} region (see Section~\ref{sec:gas}). 
It seems that the expanding S305 H\,{\sc ii} region is sweeping up surrounding material. 
Our multi-wavelength observational results favour the positive feedback of two massive O-type stars in S305. 
Hence, in the S305 H\,{\sc ii} region, the ``collect and collapse" process for sequential star formation might be applicable.

\begin{figure}
\epsscale{1.0}
\plotone{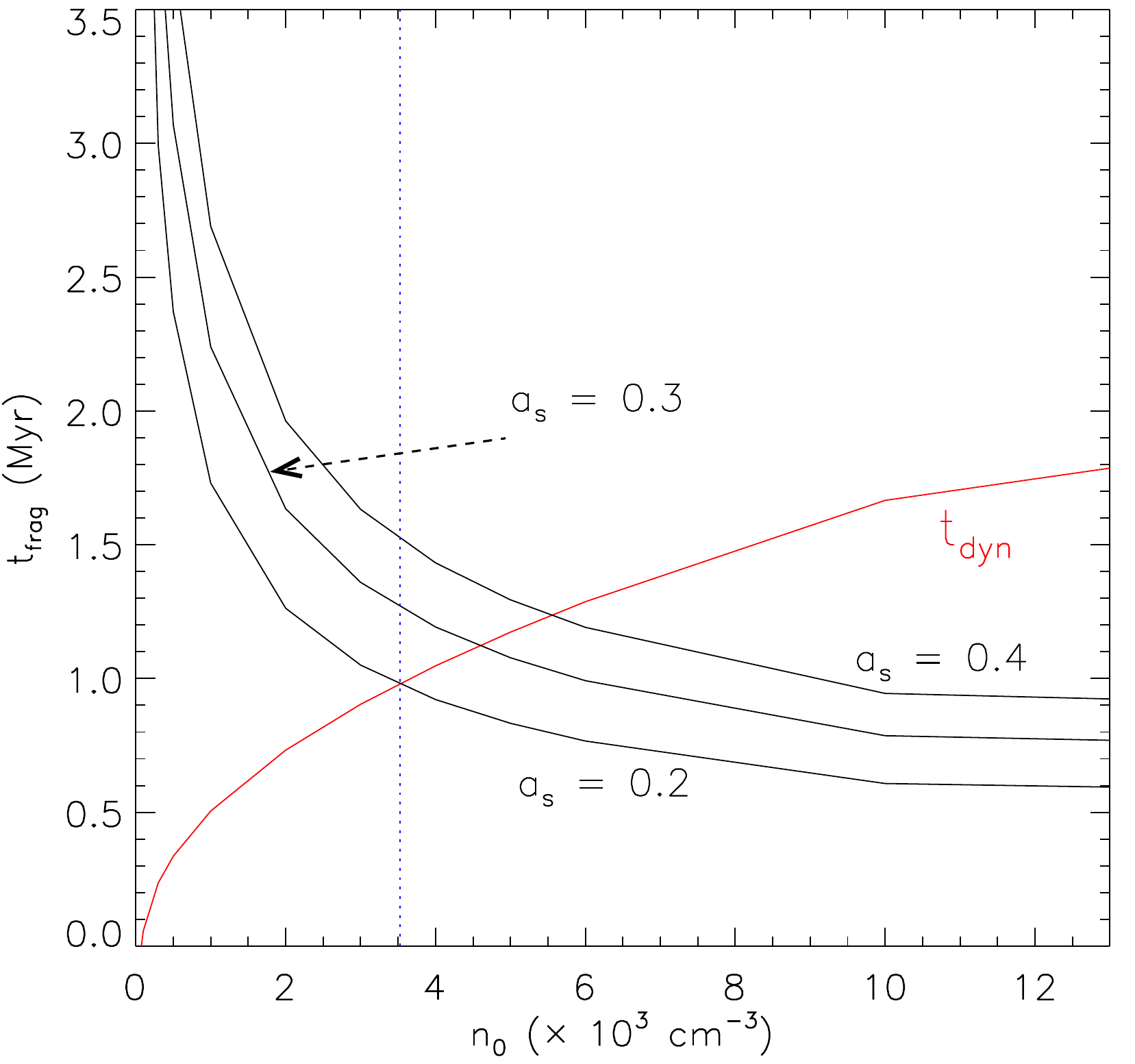}
\caption{Dynamical time (red curve) and fragmentation timescale (black curves) as a function of the initial density of the ambient medium. The
fragmentation timescale is computed for three values of sound speeds of neutral gas ($a_{s}$ = 0.2 , 0.3, and 0.4~km~s$^{-1}$). A dotted vertical line is shown at density of 3530~cm$^{-3}$.}
\label{tfig6}
\end{figure}

To explore the onset of the ``collect-and-collapse" process, one has to calculate the dynamical age ($t_\mathrm{dyn}$) of 
the H\,{\sc ii} region (see equation~2 in this paper) and the fragmentation time scale ($t_\mathrm{frag}$). 
Following the theoretical model of \citet{whitworth94}, one can compute the timescale at which fragmentation 
and collapse will onset in a swept-up shell around an expanding H\,{\sc ii} region \citep[see also equation~13 in][]{xu17}.
More details of similar analysis can also be found in \citet{dewangan12} \citep[see also][]{dewangan16}. 
Figure~\ref{tfig6} displays the plot between $t_\mathrm{frag}$ and $t_\mathrm{dyn}$ as a function of the initial ambient
density ($n_\mathrm{0}$). We have also computed $t_\mathrm{frag}$ for different values of sound speed ($a_{s}$) = 0.2, 0.3, and 0.4 km s$^{-1}$. 
Figure~\ref{tfig6} shows that the dynamical age of the H\,{\sc ii} region exceeds the fragmentation time scale 
for $n_\mathrm{0}$ = 3530~cm$^{-3}$ (at $a_{s}$ = 0.2). If we assume the density $n_\mathrm{0}$ = 1000 cm$^{-3}$ of S305 then the 
``collect and collapse" process may not be applicable in S305. If we consider the value of $n_\mathrm{0}$ $>$ 3530~cm$^{-3}$ then  
the ``collect and collapse" triggered star formation process might have occurred in S305, which seems possible in the S305 H\,{\sc ii} region. 
Earlier, in some Galactic H\,{\sc ii} regions such as Sh 2-104 \citep{deharveng03,xu17}, RCW 79 \citep{zavagno06}, 
Sh 2-219 \citep{deharveng06}, RCW 120 \citep{zavagno07,zavagno10}, Sh 2-212 \citep{deharveng08}, 
Sh 2-217 \citep{brand11}, and Sh 2-235 \citep{dewangan16}, the ``collect and collapse'' mechanism 
has been observationally examined. 

A careful analysis of our new GMRT maps has enabled us to peer into the ``ionized shell", which is investigated in the 
direction of the horseshoe envelope. 
We do not detect any dust and CO emission toward the ionized shell or around the positions of the two massive O-type 
stars. The negative feedback of massive O-type stars seems to explain the absence or destruction or dissipation of molecular materials toward the ionized shell. 
The GMRT radio continuum maps at 610 and 1280 MHz reveal the presence 
of overdensities of the ionized emission (see Section~\ref{subsec:radio1}). 
Molecular condensations, PAH emission, dust clumps, and H$_{2}$ emission are also traced toward the horseshoe envelope surrounding the ionized shell, where noticeable YSOs are also found. 
The existence of the horseshoe envelope and the ionized shell might have been produced by the positive feedback of two massive O-type stars.
\subsection{Non-thermal emission in S305} 
\label{sec:disc_nonthermal}
Thermally emitting sources have a spectral index between $-0.1$ (optically thin plasma) and 2 (optically thick plasma). On the other hand, non-thermal radio sources emit synchrotron radiation produced by relativistic electrons and are often traced with $\alpha < -0.25$. Many Galactic sources, such as supernova remnants and massive colliding-wind binaries, display non-thermal emission with $\alpha \approx -0.5$, while extragalactic objects generally show a steeper $\alpha \approx -1$ \citep[e.g.,][]{rybicki79,longair92,bihr16}. Thus, synchrotron emission is often associated with astrophysical shocks. In an H\,{\sc ii} region, the intense energetic feedback of a massive OB star (i.e., stellar wind, ionized emission, and radiation pressure) can produce shocked regions. In the case of O-type stars, the pressure due to the ionized gas dominates over the wind pressure and radiation pressure \citep[see Eqs.~10--12 in][]{pandey20}.  
The detection of non-thermal emission in H\,{\sc ii} regions indicates the presence of a population of relativistic electrons \citep[e.g.,][]{nandakumar16,veena16}. \citet{padovani19} explained the observed non-thermal emission in H\,{\sc ii} regions as synchrotron radiation from locally accelerated electrons --probably in shocked regions of gas-- restrained in a magnetic field. 

The spectral index map shown in Figure~\ref{fig6}a reveals the existence of several sub-regions with $\alpha < -0.6$ in the S305 H\,{\sc ii} region. This suggests the presence of relativistic particles accelerated in a shock with a low Mach number. Such hypothesis is consistent with the expectation of a slow shock with velocity $v_\mathrm{sh} \sim 30$~km~s$^{-1}$ moving in a medium with a sound speed of $c_\mathrm{s} \sim 10$~km~s$^{-1}$, which corresponds to a Mach number $M \sim 3$. 
The compression factor $r$ of the shock determines the spectral index of the accelerated particle energy distribution, $N(E) \propto E^{-p}$, as $p = (r+2)/(r-1)$. 
For a strong shock, $r=4$ and the canonical value $p=2$ is obtained, which yields a radio spectral index of $\alpha = -0.5$; instead, for $M \sim 3$ we get $r \sim 3$, which yields $p \sim 2.5$ and $\alpha \sim -0.75$. 
In this context, regions with more negative spectral indices ($\alpha < -0.6$) can hint the presence of weak shocks. Another possibility for having a soft radio spectral index is that electrons cool down locally by processes such as inverse-Compton scattering or synchrotron, but this seems unlikely given the ambient conditions (diluted radiation fields and no particularly strong magnetic fields, respectively). Instead, the cooling of electrons is dominated by Bremsstrahlung losses which are efficient given the high ambient density; these losses do not soften the electron energy distribution.

One of aspects to take into account is the assumption of a constant spectral index for 
the non-thermal component.~Absorption/suppression processes have a larger effect at low radio frequencies, which can harden (i.e., make less negative) the spectral index at lower frequencies. We suggest that such behavior can be produced in H\,{\sc ii} regions by the Razin-Tsytovich effect, which suppresses the synchrotron emission below a frequency $\nu_\mathrm{RT}[\mathrm{Hz}] = 20 n_\mathrm{e}[\mathrm{cm}^{-3}]/B[\rm{G}]$ \citep{melrose80}.

\begin{figure}
\epsscale{1.0}
\plotone{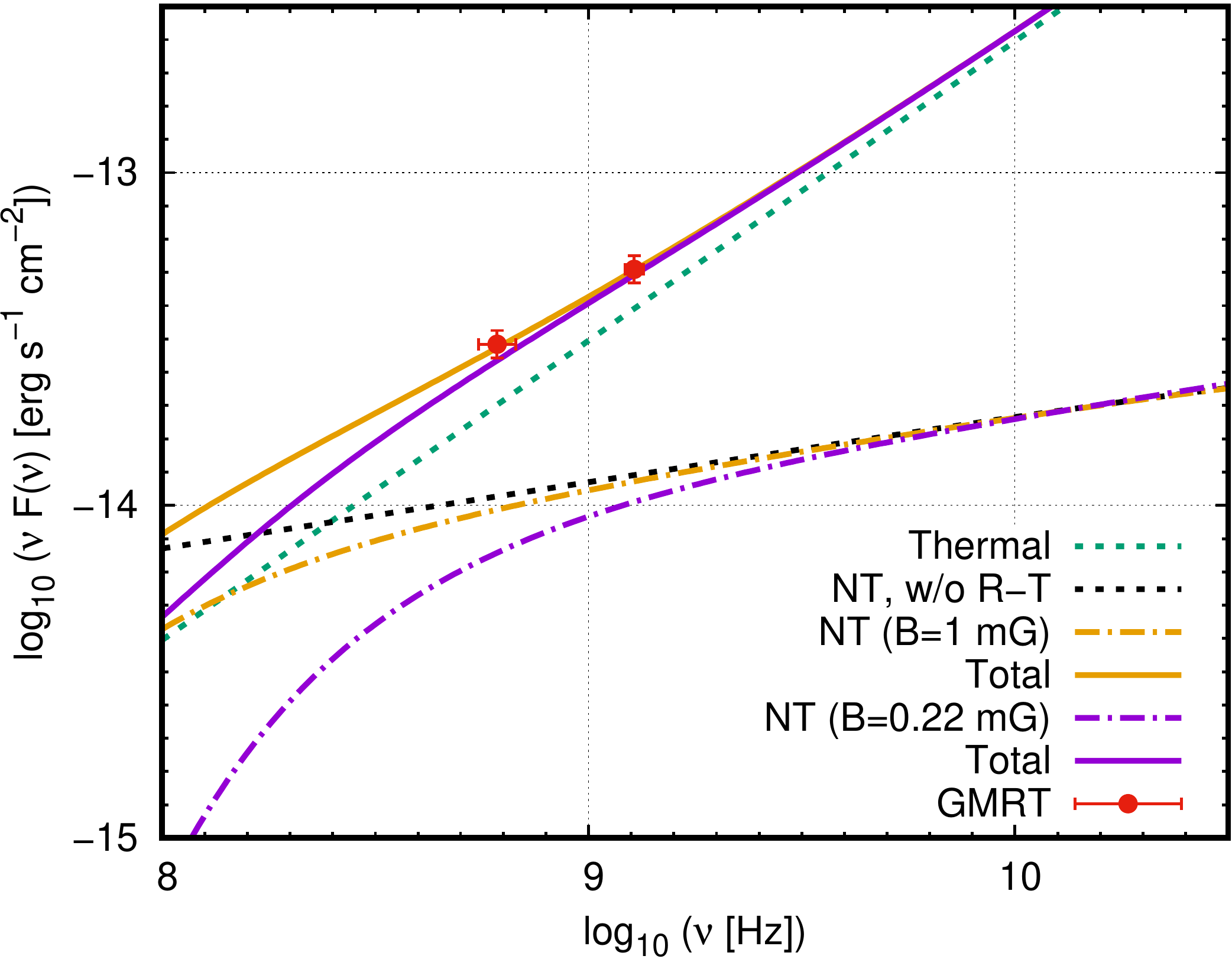}
\caption{Observed spectral energy distribution of S305 at radio frequencies. 
We fit the spectrum as a composite spectrum of a thermal component from the ionized gas plus a non-thermal (NT) component from locally-accelerated relativistic electrons. We show the emission model for extreme scenarios of a 
high magnetic field ($B \approx 1$~mG) and a low magnetic field ($B \approx 0.22$~mG). For comparison, we also show the synchrotron component without accounting for the Razin-Tsytovich effect (dotted black line; see Sect.~\ref{sec:disc_nonthermal} for details).}
\label{ttfig6}
\end{figure}

This effect has been studied in different astrophysical contexts such as YSOs \citep{freeney19}, supernova remnants \citep{fransson98}, massive colliding-wind binaries \citep{dougherty03}, and compact $\gamma$-ray binaries \citep{marcote15}. In Figure~\ref{ttfig6}, we show how this effect makes the non-thermal emission less intense at low frequencies than a power-law extrapolation from high frequencies. This effect can be used to constrain the magnetic field intensity in the 
S305 H\,{\sc ii} region. The negative spectral index observed requires that $\nu_\mathrm{RT} < 300$~MHz; for a value of $n_e \sim 5\times10^3$~cm$^{-3}$, this translates into $B > 0.2$~mG. Moreover, we can set an upper limit to the magnetic field of $B < 1$~mG given that the shock velocity becomes sub-alfvenic for higher values \citep[e.g.,][]{padovani19}.

The presence of regions with $\alpha > -0.5$ could be due to either a more intense thermal emission there or due to a stronger suppression of the low-frequency non-thermal emission. The harder spectral indices (i.e., near the rims) can be attributed to a higher particle number density in them, consistent with either scenario. The main difference between the two scenarios is that if it is due to thermal emission being optically thicker we expect to see a harder spectrum at higher frequencies, whereas if it is due to suppression of non-thermal emission we expect a harder spectrum at lower frequencies. This can be seen in Figure~\ref{ttfig6}.

We can learn more about the relativistic particle population by fitting the observed spectrum with a non-thermal emission model. We follow a similar approach as in \citet{prajapati19} and model the non-thermal electron distribution as a power law with a spectral index $p = 2.6$, with a hardening at $E_\mathrm{e} < 10$~MeV due to ionization losses, and a high-energy cutoff at $\sim 100$~GeV produced by synchrotron losses. The particle distribution normalization is set by the condition $U_\mathrm{NT} = \eta_\mathrm{mag} U_\mathrm{mag}$, with $U_\mathrm{NT}$ the energy density in relativistic particles and $\eta_\mathrm{mag}$ is a parameter in the range $3\times10^{-3}$--0.75 (in accordance to the allowed range for $B$). We also assume that the ratio between the energy density between protons and electrons is $K_\mathrm{e,p} = 0.01$. 

We obtain that fitting the observed radio flux requires $U_\mathrm{NT} \sim 60$--900~eV~cm$^{-3}$ (the larger values correspond with larger $\eta_\mathrm{mag}$). 
This value is much greater than the Galactic cosmic ray energy density, which is $U_\mathrm{GCR} \sim 1$~eV~cm$^{-3}$. Thus, ambient cosmic rays are not sufficient to explain the observed synchrotron flux. This supports the hypothesis that the cosmic rays at S305 are accelerated locally, as consistent with the scenario proposed by \citet{padovani19}.
%
\section{Summary and Conclusions}
\label{sec:conc}
%
To uncover physical processes in the S305 H\,{\sc ii} region, we have carried out an analysis of multi-scale and multi-wavelength data of a field (size $\sim$18$'$.7 $\times$ 18$'$.7) containing S305, which includes new high angular resolution radio continuum maps at 610 and 1280 MHz observed using the GMRT facility. Various observational data sets allow us to disentangle the multi-phase structures of S305. 
The main results of our analysis are:\\
$\bullet$ An extended IR shell (extension $\sim$7.5 pc) is identified as the largest structure in the S305. 
This sphere-like shell encloses the radio continuum emission.\\ 
$\bullet$ The S305 H\,{\sc ii} region (extension $\sim$5.5 pc) is known to be excited by massive O8.5V (VM4) and O9.5V (VM2) stars.
This result is in agreement with the analysis of the GMRT radio maps at 610 and 1280 MHz.
The dynamical age of the S305 H\,{\sc ii} region is estimated to be $\sim$1.7 Myr for n$_\mathrm{0}$ = 10$^{4}$ cm$^{-3}$.\\
$\bullet$ The molecular gas in S305 is examined in a velocity range of [39.65, 48.1] km s$^{-1}$. 
Its study provides a signature of an expanding shell of molecular gas in S305.\\
$\bullet$ Regularly spaced molecular condensations and dust clumps are traced toward the edges of the IR shell, 
where PAH and H$_{2}$ emission is also detected. These outcomes provide an evidence of collected material (molecular and dust) 
along the IR shell around the S305 H\,{\sc ii} region. Noticeable YSOs are also seen toward some of the clumps.\\
$\bullet$ The dynamical age of the 
S305 H\,{\sc ii} region is longer than the fragmentation time scale of accumulated gas layers in S305. \\
$\bullet$ An ionized shell (extension $\sim$2.15 pc) is traced in the H$\alpha$ image around 
the positions of two massive O-type stars, where a molecular gas deficient region/cavity is investigated. 
The cavity is also found to be surrounded by a horseshoe envelope (extension $\sim$2.3 pc), which 
is associated with the PAH, H$_{2}$, molecular, and dust emission. Noticeable YSOs are also detected toward the horseshoe envelope.\\ 
$\bullet$ The {\it Herschel} images show that the observed structures in S305 are associated with the warm dust emission at T$_\mathrm{d}$ = 17.5--27~K. 
A signature of the existence of an obscured filament is found in S305, which shows 
a relatively cold dust emission at T$_\mathrm{d}$ = 13.5--14.2~K. 
The IR shell containing two massive O-type stars appears to be located at the center of the filament.\\ 
$\bullet$ The analysis of molecular line data suggests that the filament appears to be located in the front of both the IR shell and horseshoe envelope. 
Furthermore, the edges of the IR shell are found to be located in the front of the horseshoe envelope.\\ 
$\bullet$ Non-thermal radio emission in S305 is observed with an average spectral index of $\sim-$0.45. 
The variations in $\alpha$ range from $-1.1$ to 1.3. We interpret that this spectrum corresponds to a combination of soft synchrotron emission and thermal emission from the ionized gas. We show that suppression of low-frequency emission 
by the Razin-Tsytovich effect can be relevant in the context of H\,{\sc ii} regions. 
We estimated the magnetic field in the H\,{\sc ii} region to be $\sim$0.2--1~mG. The high content of relativistic particles needed to produce the observed emission requires them to be accelerated {\it in situ}. Finally, the soft radio emission is consistent with 
a low Mach number shock with $V_\mathrm{sh} \sim 30$~km~s$^{-1}$. Future observations exploring the unobserved frequency ranges below 600~MHz and above 1.4~GHz are expected to further constrain these values.
\smallskip \smallskip  

The overall findings show the observational evidence of positive feedback of O-type stars in S305, which appears to explain the observed morphology in S305. 

\acknowledgments  
We thank the anonymous reviewer for several useful comments and 
suggestions, which greatly improved the scientific contents of the paper.  
The research work at Physical Research Laboratory is funded by the Department of Space, Government of India. 
We thank the staff of the GMRT, who made these observations possible. 
GMRT is run by the National Centre for Radio Astrophysics of the Tata Institute of Fundamental Research.  
This work is based [in part] on observations made with the {\it Spitzer} Space Telescope, which is operated by the Jet Propulsion Laboratory, California Institute of Technology under a contract with NASA. 
This publication makes use of data from FUGIN, FOREST Unbiased Galactic plane Imaging survey with the Nobeyama 45-m telescope, a legacy project in the Nobeyama 45-m radio telescope. 
DKO acknowledges the support of the Department of Atomic Energy, Government of India, under project NO. 12-R\&D-TFR-5.02-0200.
TB is supported by the National Key Research and Development Program of China through grant 2017YFA0402702. TB also acknowledges support from the China Postdoctoral Science Foundation through grant 2018M631241. We thank F. Navarete for providing the narrow-band H$_{2}$ image through the survey of extended H$_{2}$ emission from massive YSOs.
%

%

%
\end{document}